\documentclass[aps,prb,reprint,longbibliography]{revtex4-2}
\usepackage[utf8]{inputenc}
\usepackage{amssymb}
\usepackage{amsmath}
\usepackage{graphicx}
\usepackage[caption=false,font=small,subrefformat=parens,labelformat=parens]{subfig}
\usepackage[colorlinks=true,linkcolor=blue,allcolors=blue]{hyperref}
\usepackage{nicefrac}
\usepackage{float}
\usepackage{comment}
\newcommand\numberthis{\addtocounter{equation}{1}\tag{\theequation}}
\usepackage{physics}
\newcommand{\mbfk}{\mathbf{k}}
\newcommand{\mbfp}{\mathbf{p}}
\newcommand{\mbfq}{\mathbf{q}}
\newcommand{\mbfK}{\mathbf{K}}

\newcommand{\expect}[1]{\left\langle #1 \right\rangle}
\usepackage{siunitx}
\usepackage{lipsum}

\usepackage[table]{xcolor}

\usepackage{subfiles}

\begin{document}

\title{Valley polarization, magnetization, and superconductivity in bilayer graphene near the van Hove singularity}
\author{Alex Friedlan}
\author{Heqiu Li}
\author{Hae-Young Kee}
\email{hy.kee@utoronto.ca}
\affiliation{Department of Physics and Centre for Quantum Materials, University of Toronto, 60 St. George Street, Toronto, Ontario, Canada, M5S 1A7}

\begin{abstract}
   The discovery of Mott insulators and superconductivity in twisted bilayer graphene has ignited intensive research into strong correlation effects in other stacking geometries. Bernal-stacked bilayer graphene (BBG), when subjected to a perpendicular electric field, exhibits phase transitions to a variety of broken-symmetry states. Notably, superconductivity emerges when BBG is in proximity to a heavy transition-metal dichalcogenide, highlighting the role of spin-orbit coupling (SOC). Here we investigate the origin of Ising SOC and its role in the competition between superconductivity and spin- and valley-polarized states in BBG. Starting from strong electron-electron interactions on the BBG lattice, we derive a low-energy effective model near the valleys that incorporates both density-density and spin-spin interactions. Using self-consistent mean-field theory, we map out the BBG phase diagram. Our findings reveal that near the van Hove filling, a mixed spin- and valley-polarized phase dominates over superconductivity. Away from the van Hove filling, a spin-polarized, spin-triplet superconducting state arises, characterized by an in-plane orientation of the magnetic moment and an out-of-plane orientation of the $d$-vector. Contrary to previous proposals, we find that Ising SOC favours spin-valley order while suppressing superconductivity near the van Hove singularity. We discuss other potential proximity effects and suggest directions for future studies.
\end{abstract}
\maketitle

\section{Introduction}\label{Intro}

Twisted bilayer graphene has emerged as a popular platform for investigating strongly-correlated electronic states, primarily due to the appearance of flat Moiré bands at certain twist angles that amplify electron-electron interactions \cite{twistedunconventionalSC,Bistritzer2011,Andrei2020}. Prior to this wave of interest, Bernal-stacked bilayer graphene (BBG) was recognized for its strong electronic correlations at low energies \cite{RozhkovReview2016}. Several candidate phases were proposed, including both gapped and gapless states \cite{Weitz2010,Bao2012,Mayorov2011,Martin2010}. Until recently, however, superconductivity in BBG has remained largely unexplored.

Renewed interest in BBG was sparked by a pair of 2021 papers that observed a sequence of unusual symmetry-breaking electronic states as the bias field and electron density were tuned \cite{isospin,cascade}. Notably, when the BBG was subjected to an in-plane magnetic field, a region of superconductivity emerged within the phase diagram, pointing to unconventional superconductivity. A subsequent study found a similar enhancement of the superconducting state when the BBG was placed on a substrate of monolayer tungsten diselenide (WSe\textsubscript{2}) \cite{WSe2}. Due to the low atomic mass of carbon, graphene exhibits intrinsically tiny spin-orbit coupling (SOC) \cite{Min2006,Konschuh2010,Konschuh2012,Sichau2019,David2019}. However, when placed on a substrate with a heavy atom (i.e. tungsten), graphene experiences a proximity-induced SOC enhancement of up to two orders of magnitude \cite{proximityDFT,WangZ2015,WangZ2016,WangD2019,Szentpeteri2024}. SOC is believed to play a key role in the enhancement of the superconducting state \cite{WSe2,ZhangY2024}.

Recent experiments have further revealed the complex structure of the BBG phase diagram. In addition to superconductivity, spin- and valley-polarized phases \cite{Weitz2022,Holleis2023,SC_SOC_July2024,BLGSC_Review}, as well as exotic Wigner crystal and fractional quantum Hall states \cite{Seiler2024ID,HuangK2022} have been reported. One study has even identified two separate superconducting regimes, one of which is argued to arise from a nematic normal state characterized by broken rotational symmetry \cite{Holleis2023}. Rhombehedral-stacked multilayer graphene, which shares many similarities with BBG, is also being vigorously explored \cite{ZhouH2021Rhombo,RTG_Hunds,Arp2023,RTG_penta,Patterson2024,ZhouW2024}.

Theoretical efforts to understand the various normal-state phases in BBG under a displacement field have also been undertaken \cite{hartreefock1,hartreefock2,hartreefock3}. These studies examine the normal-state phase diagrams in the presence of long-range Coulomb interactions \cite{hartreefock2,hartreefock3} and short-range Hund's coupling \cite{hartreefock1}, but do not address superconductivity. In addition, some work has focused on generic two-dimensional systems with spin and valley degrees of freedom \cite{Raines2024_simpler,LeeYC2024}.

Several proposals have been put forward to explain the origin of superconductivity in the context of Ising SOC. A recent study proposed a mangon exchange mechanism for superconductivity arising from a spin-canted normal state, favouring $s$-wave pairing \cite{DongZ2024}. Another study has suggested combined $p$- and $d$-wave superconductivity that emerges from competition with an intravalley current density wave connecting the small Fermi pockets near the van Hove singularity \cite{SonJH2024}. $f$-wave superconductivity is also being explored \cite{Curtis2023,Jimeno-Pozo2023}, and has been argued to be the most favourable superconducting channel \cite{ChouYZ2021,ChouYZ2022b}. Clearly, the symmetry and origin of the superconducting state in BBG remain open questions requiring further exploration (see also Refs. \cite{LiZ2023,Wagner2023,Ghazaryan2023,Koh2024,LuDC2022,Gindikin2023}).

In this work, we examine the phase diagram of BBG near electron densities corresponding to the van Hove singularity (vHS) to explore the interplay of broken-symmetry phases. Our interacting model is based on the idea that, at low energies, the biased BBG lattice resmebles a honeycomb lattice with a staggered sublattice potential, as studied in Ref. \cite{spintriplet}. By applying a Schrieffer-Wolff transformation and incorporating on-site Hubbard, nearest-neighbour (NN), and next-nearest-neighbour (NNN) repulsive interactions, we obtain effective intervalley density-density and spin-spin interactions. The resulting effective Hamiltonian yields in valley-polarized, spin-polarized, and spin-triplet superconducting states. We perform self-consistent mean-field (MF) calculations with and without SOC to investigate the role of SOC in promoting superconductivity.

This paper is organized as follows. In Sec. \ref{Theory}, we review the single-particle tight-binding Hamiltonian and examine the Fermi surface near half-filling. We then derive the proximity-induced SOC from the adjacent WSe\textsubscript{2} layer, starting from the atomic SOC in the tungsten $d$ orbitals. In Sec. \ref{Interacting}, we introduce the interacting Hamiltonian on the BBG lattice, including electron-electron interactions up to next-nearest-neighbour distance. Performing a Schrieffer-Wolff transformation, we obtain the effective interacting Hamiltonian for electrons near the valleys, which takes the form of intervalley density-density and spin-spin interactions, akin to Hund's coupling in valley degrees of freedom. After introducing the MF order parameters in Sec. \ref{MFOP}, we present the self-consistent MF phase diagram and results for a few representative parameter sets in Sec. \ref{MF}. Finally, in Sec. \ref{disco}, we summarize our findings, discuss the implications, and outline open questions for future studies.
 
\section{Single-particle Hamiltonian}\label{Theory}

\subsection{Tight-binding model}\label{TightBindingSec}
\begin{figure}[]
    \centering
    \includegraphics[width=0.9\linewidth]{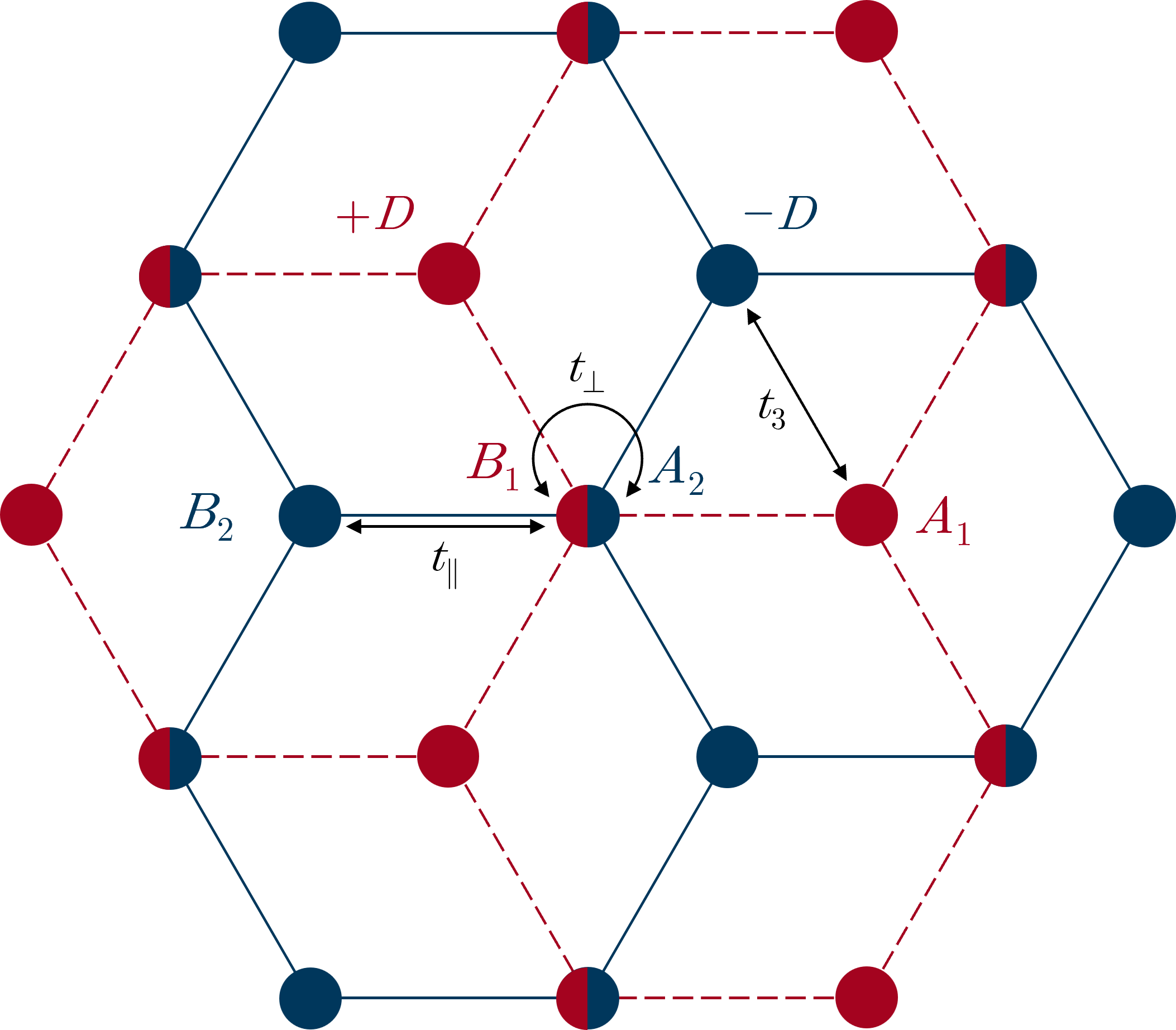}
    \caption{Bernal bilayer graphene lattice. The $A_1$ and $B_1$ atoms (red) are in the top layer ($l=1$), while the $A_2$ and $B_2$ atoms (blue) are in the bottom layer ($l=2$). The dominant hopping processes $t_{\parallel}$, $t_{\perp},$ and $t_3$ in Eq. (\ref{hkmatrixx}) are indicated. Due to strong $t_{\perp}$ between overlapping $A_2$ and $B_1$ sites, $A_1$ and $B_2$ orbitals form the low-energy bands near half-filling.}
    \label{BLG lattice}
\end{figure}

We begin with a short review on the tight-binding model of the BBG lattice. We consider bilayer graphene in an AB (Bernal) stacking arrangement as shown in Fig. \ref{BLG lattice}. We include the NN intralayer hopping $t_\|$, the interlayer dimer hopping $t_\perp$, and the non-dimer interlayer hopping $t_3.$ To construct the tight-binding model, we introduce the creation (annihilation) operators $a_{l,\mbfk}^\dagger$ and $b_{l,\mbfk}^\dagger$ ($a_{l,\mbfk}$ and $b_{l,\mbfk}),$ corresponding to sublattices $A$ and $B,$ respectively. The layers are indexed by $l \in\left\{ 1,2\right\}$ and the associated momentum is denoted by $\mbfk.$ In addition to the hopping Hamiltonian, an external electric displacement field $D$ is introduced perpendicular to the plane of the graphene, leading to the potential difference term $\mathcal{H}_D = D (n_1 - n_2),$ where $n_l$ is the density operator for layer $l$.

Introducing the spinor ${{\tilde \Psi}_\mbfk = \left(a_{1,\mbfk},b_{1,\mbfk},a_{2,\mbfk},b_{2,\mbfk}\right)^T,}$ the Hamiltonian can be written $\mathcal{H}_0 = \sum_\mbfk {\tilde \Psi}_\mbfk^\dagger H_0(\mbfk){\tilde \Psi}_\mbfk,$ where $H_0(\mbfk)$ takes the explicit matrix form \cite{McCannKoshinoBLG,McCannTB,mypaper}
\begin{equation}\label{hkmatrixx}
     H_0(\mbfk) = \\ \begin{pmatrix}
    D & t_\|f^*_\mbfk & 0 & t_3 g^*_\mbfk \\
    t_\|f_\mbfk & D & t_\perp & 0 \\
     0 & t_\perp & -D & t_\|f^*_\mbfk \\
    t_3 g_\mbfk & 0 & t_\|f_\mbfk & -D
    \end{pmatrix},
\end{equation}
where $f_\mbfk = 1+e^{-i\mbfk\cdot\mathbf{a}_1}+e^{-i\mbfk\cdot\mathbf{a}_2}$ and $g_\mbfk =  e^{-i\mbfk\cdot\mathbf{a}_1}+e^{-i\mbfk\cdot\mathbf{a}_2} +e^{-i\mbfk\cdot(\mathbf{a}_1+\mathbf{a}_2)}.$ Here, $\mathbf{a}_1 = a_0(3\hat{\mathbf{x}}+\sqrt{3}\hat{\mathbf{y}})/2$ and $\mathbf{a}_2 = a_0(3\hat{\mathbf{x}}-\sqrt{3}\hat{\mathbf{y}})/2$ are the primitive translation vectors. Typical strengths for the hopping parameters are $t_\|=-3.3$ eV, $t_\perp=0.42$ eV, and $t_3=-0.315$ eV \cite{hopping,McCannTB}, which will be used throughout the paper. The eigenvalues are
\begin{equation}\label{valence}
    \pm E^{\pm}_\mbfk = \pm \sqrt{\left(t_\|^2+\frac{t_3^2}{2}\right)|f_\mbfk|^2+D^2+\frac{t_\perp^2}{2}\pm\epsilon_\mbfk^2}\,,
\end{equation}
where
\begin{align*}\label{epsilon}
    \epsilon_\mbfk & =\left[\left(t_\|^2+\frac{t_3^2}{4}\right)|f_\mbfk|^4t_3^2 +\left(\left(4D^2+t_\perp^2\right)t_\|^2-\frac{t_3^2 t_\perp^2}{2}\right)|f_\mbfk|^2 \right. \\ & \left. 
    + t_\perp t_3t_\|^2\left(g^*_\mbfk f^2_\mbfk + g_\mbfk \left(f^*_\mbfk  \right)^2\right) + \frac{t_\perp^4}{4}\right]^{1/4}. \numberthis
\end{align*}

\begin{figure}
    \centering
    \includegraphics[width =0.9 \linewidth]{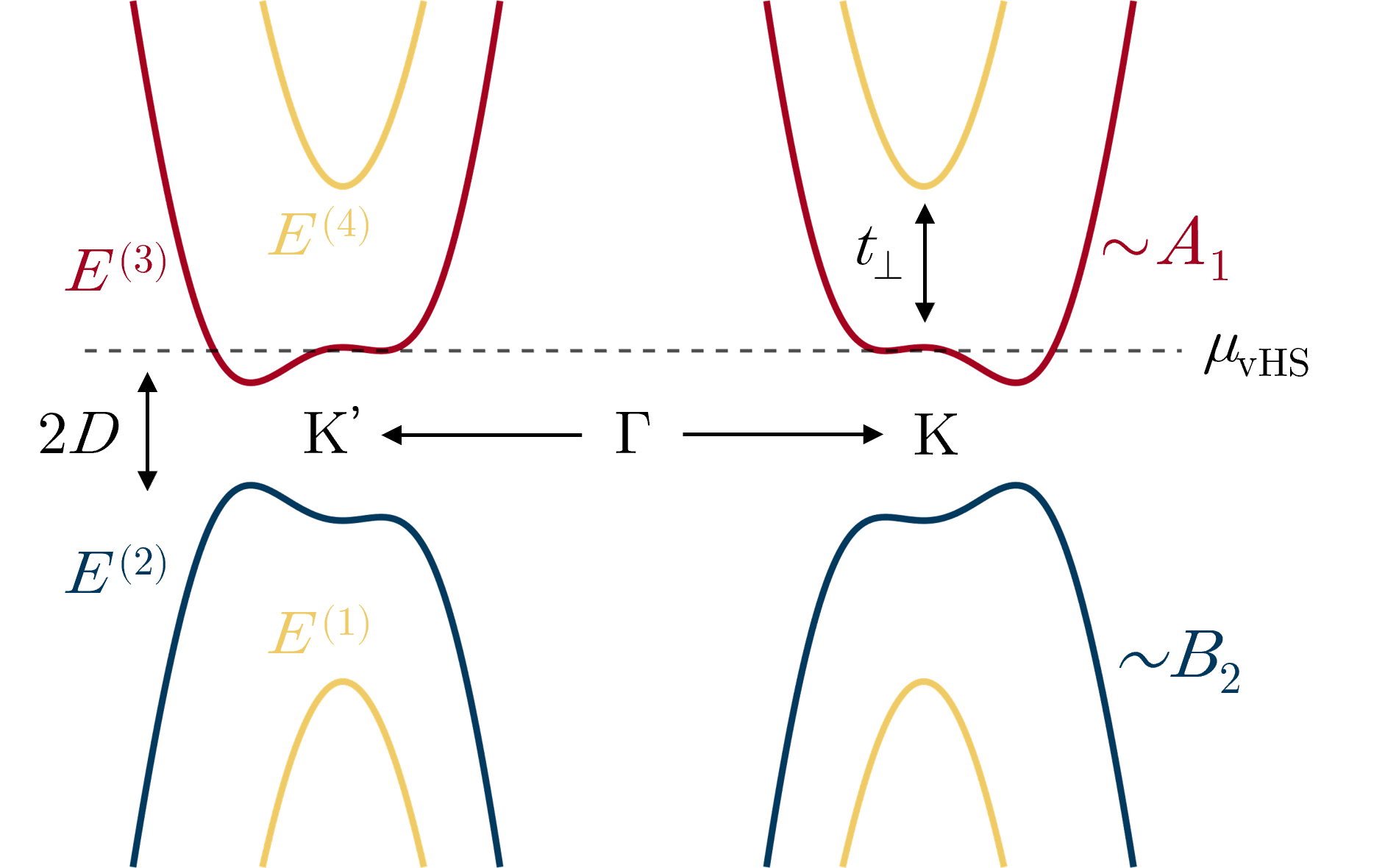}
    \caption{Low-energy band structure of biased BBG near $K$ and $K'.$ The red and blue bands ($E^{(3)}$ and $E^{(2)}$) originate predominantly from the $A_1$ and $B_2$ orbitals, respectively. These bands are gapped out by the displacement field $D.$ The yellow bands ($E^{(4)}$ and $E^{(1)}$) originate from symmetric and antisymmetric combinations of the $A_2$ and $B_1$ dimer orbitals, and are pushed away from charge neutrality by the interlayer hopping $t_\perp$ (not to scale). The chemical potential $\mu$ is shown tuned to the van Hove singularity within $E^{(3)}.$}
    \label{bands}
\end{figure}

The broken-symmetry phases arise near half-filling in the biased BBG system. In this regime, the low-energy bands are located near the Brillouin-zone corners at $\mbfK$ and $\mbfK'.$ In our coordinate system, $\mbfK$ and $\mbfK'=-\mbfK$ are found along $k_x=0$ at $\mbfk = (0,\pm 4\pi/3\sqrt{3}a_0).$ Fig. \ref{bands} displays the four electronic bands $E^{(i)},$ with $i=1,..,4$ labelled in order of increasing energy about the $K$ points. The displacement field $D$ gaps out the electronic spectrum, resulting in extremely flat bands at certain field strengths \cite{Asymmetry,EFE,directobs,controlling}. The displacement field also polarizes the electronic wavefunctions to just one of the two BBG layers, allowing for tunable proximity effects \cite{Khoo2017,Island2019}.

The outermost bands $E^{(4)}$ and $E^{(1)}$ with energy $\pm E^{+}_\mbfk$ arise from symmetric and antisymmetric combinations of the overlapping $A_2$ and $B_1$ dimer sites, which are hybridized by $t_\perp$ and pushed away from charge neutrality, as shown in Fig. \ref{bands}. Thus, the low-energy physics are dominated by the innermost bands $E^{(3)}$ and $E^{(2)},$ which have energy $ \pm E^{-}_\mbfk,$ respectively. As has been observed previously \cite{szaboroy1,ChouYZ2022}, these bands originate primarily from the isolated (non-dimer) $A_1$ and $B_2$ orbitals. Specifically, the $E^{(2)}$ band (blue) is composed almost exclusively of contributions from the $B_2$ orbitals. Similarly, the $E^{(3)}$ band (red) is composed primarily of contributions from the $A_1$ orbitals. Roughly speaking, if we denote by $c_\mbfk$ the eigenvector associated with the $E^{(3)}$ band, $c_\mbfk \sim \alpha_{\bf k} a_{1,\mbfk} + \beta_\mbfk b_{2,\mbfk},$ where $|\alpha_\mbfk|\rightarrow 1$ and $|\beta_{\bf k}| \rightarrow 0$ as $\mbfk\rightarrow\pm\mbfK.$ A similar parameterization holds for the $E^{(2)}$ band. In this way, excess charges above neutrality are localized to layer $l=1$ \cite{Asymmetry,McCannKoshinoBLG}.

Although we use the full dispersion for the purposes of our mean-field calculations, it is instructive to examine the Taylor expansion of the low-energy bands in the vicinity of $\mbfK$ and $\mbfK'.$ To fourth order in $\mbfk,$
\begin{align*}\label{lowenergyexpansion}
    E^-_{\mbfk} & \approx D + u\left(k_x^2+k_y^2\right) + w\left(k_x^2+k_y^2\right)^2  \\& \pm v\left(k_y^2 - 3k_x^2 \right)k_y, \numberthis
\end{align*}
where the plus and minus signs correspond to the $K$ and $K'$ valleys, respectively. $E^-_{\mbfk}$ can be simplified by collecting the symmetric and antisymmetric contributions $\varepsilon_\mbfk^{s} \equiv D + u\left(k_x^2+k_y^2\right) + w\left(k_x^2+k_y^2\right)^2$ and $\varepsilon_\mbfk^{a} \equiv v\left(k_y^2 - 3k_x^2 \right)k_y.$ Introducing the valley index $\tau = \{+,-\},$ we have $E_{\mbfk}^- \approx \varepsilon_\mbfk^{s} +\tau \varepsilon_\mbfk^{a}$ near $\mbfK$ or $\mbfK'.$ The factors $u,v,$ and $w$ are constant functions of the hopping parameters and the displacement field $D,$ and their expressions are given in Appendix \ref{lownrgappendix}. 
The quartic terms are necessary to capture the main features of the full dispersion. $\varepsilon_\mbfk^s$ transforms according to the $A_1$ irreducible representation of $\mathcal{C}_{3v},$ while $\varepsilon_\mbfk^a$ transforms according to $A_2.$ The inversion symmetry is preserved in this expansion, but is now represented as valley exchange $\tau\rightarrow -\tau$ combined with $\mbfk\rightarrow-\mbfk.$ For $D/t_\perp\ll 1,$ the coefficient $v$ is proportional to the trigonal warping $t_3,$ effectively quantifying the degree of asymmetry. 

\begin{figure}
    \centering
    \includegraphics[width = 0.57\linewidth]{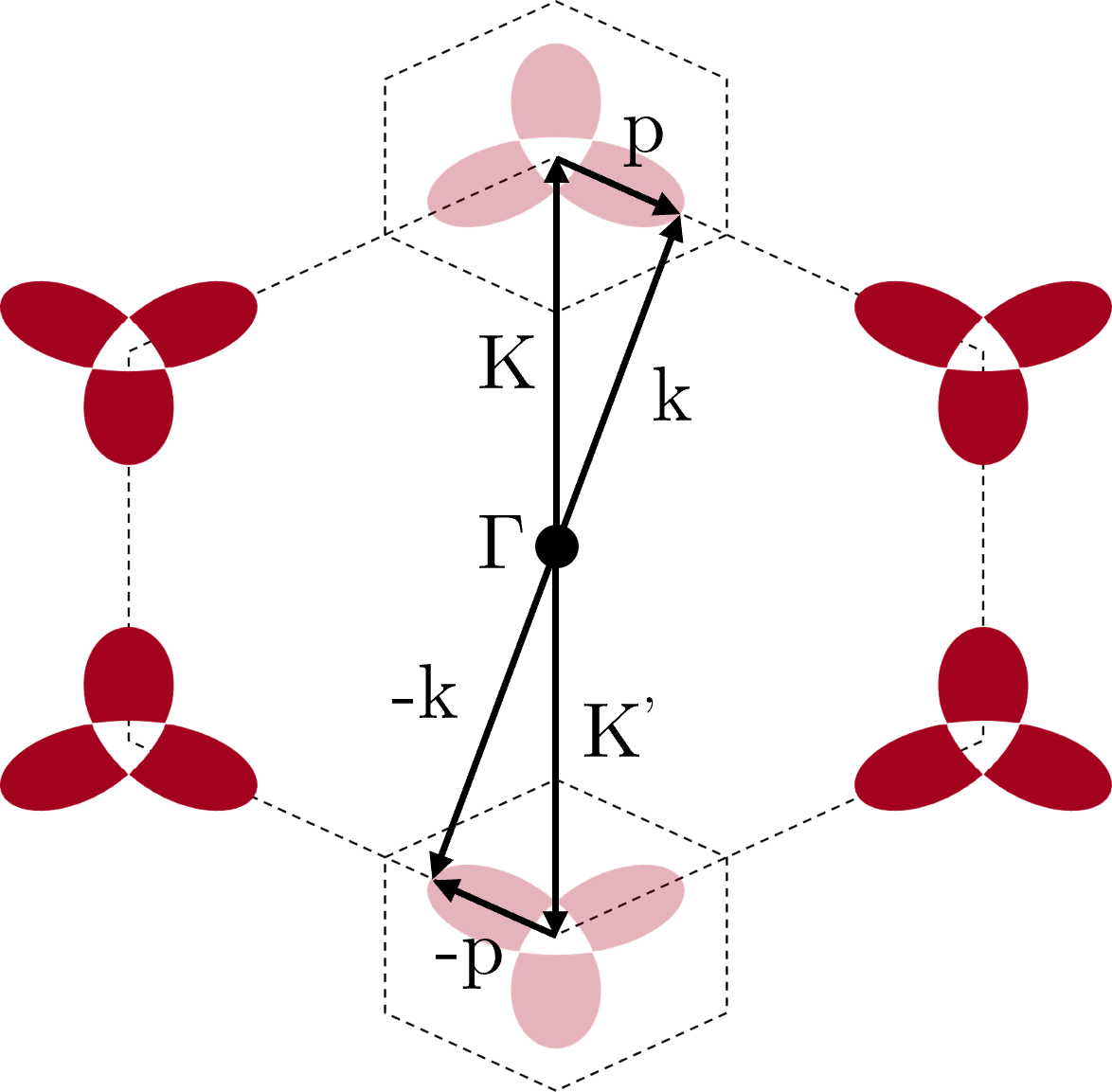}
    \caption{Low-energy patch description near the valleys. Red trefoils are low-energy Fermi surfaces at the van Hove singularity for $D=50$ meV (not to scale). The larger hexagon indicates the boundary of the first Brillouin zone. The smaller hexagons represent the momentum-space patches of our low-energy theory (also not to scale). $\mbfk$ is measured from the BZ centre, while $\mbfp=\mbfk-\mbfK$ is measured relative to the $\mbfK$ point.}
    \label{BZ}
\end{figure}

The van Hove singularities originate from the trigonal warping $t_3$ and correspond to saddle points in the electronic structure \cite{TrigWarping}. Each valley possesses three saddle points, one of which is located along the $k_x=0$ axis. The others are related by a $C_3$ rotation. At the vHS energy $\mu_{\rm vHS},$ the Fermi surface is a trefoil, as shown in Fig. \ref{BZ}. The saddle points are found at the trefoil vertices. The vHS coordinates can be computed numerically from the full dispersion Eq. (\ref{valence}) by evaluating the Hessian at the extrema, or approximately (but analytically) from the expansion in Eq. (\ref{lowenergyexpansion}) (see Appendix \ref{lownrgappendix}).

When the system is lightly electron-doped, it is reasonable to limit the single-particle theory to those bands originating from the $A_1$ sites (band $E^{(3)} \equiv E^-_{\bf k}$). Introducing the spinor ${\Psi_{\mbfp,\sigma} = (c_{+, \mbfp,\sigma},c_{-,\mbfp,\sigma})^T},$ where $\tau \in \{+,-\}$ denotes the valley, $\sigma \in \{ \uparrow, \downarrow \}$ denotes spin, and $\mbfp$ is measured from $\mbfK$ or $\mbfK'$ (see Fig. \ref{BZ}), the effective kinetic Hamiltonian near the valleys can be expressed as
\begin{equation}\label{KineticTau}
    \mathcal{H}_0 = \sum_{\mbfp, \sigma} \Psi_{\mbfp,\sigma}^\dagger \left( \xi_\mbfp \tau_0 + \varepsilon^a_{\mbfp} \tau_z \right) \Psi_{\mbfp,\sigma}.
\end{equation}
Here, $\vec \tau$ is a Pauli matrix acting in the valley degree of freedom and we have absorbed the chemical potential $\mu$ into ${\xi_\mbfp \equiv \varepsilon^s_{\mbfp}-\mu}.$ The momentum $\mbfp$ is restricted to ${|\mbfp|\ll|\mbfK|},$ as $|\mbfK|$ is the separation between the valleys.

By discarding the high-energy bands, we effectively eliminate the $A_2$ and $B_1$ dimer sites. The remaining $A_1$ and $B_2$ sites form an emergent honeycomb lattice connected by $t_3.$ This system closely resembles monolayer graphene with a staggered sublattice potential \cite{Semenoff1984,ZhouSY2007}, where here the displacement field plays the role of the staggered potential. Such models are also applicable to WSe$_2$ and other transition-metal dichalcogenides \cite{Xiao2012}.

\subsection{Proximity-induced spin-orbit coupling}\label{proximity}
We now place the BBG on a substrate of monolayer tungsten diselenide, i.e. above the top layer $l=1$. The heavy tungsten atoms introduce a sizeable atomic spin-orbit coupling term $\mathcal{H}_w=-\lambda_w\mathbf{L}\cdot\mathbf{S}$ to the tungsten $d$-orbitals of the WSe\textsubscript{2} lattice. Through a second-order hopping process, we show how electrons in the proximate graphene layer inherit Ising SOC. We do not attempt a rigorous derivation, but rather seek a qualitative understanding of its origin.

Due to the trigonal-prismatic crystal-field splitting, the $d$-orbitals separate into three levels: $d_{xz}/d_{yz},$ $d_{xy}/d_{x^2-y^2},$ and $d_{z^2}$ \cite{Gong2013,Liu2018}. We assume that the active orbitals are the $d_{xz}/d_{yz}$ level, but a similar argument holds if we consider $d_{xy}/d_{x^2-y^2}$ orbitals as the active level. In the subspace spanned by $d_{xz}$ and $d_{yz}$, $\mathbf{L}\cdot\mathbf{S}$ behaves like $L_zS_z$ and we have explicitly $L_zS_z\ket{xz,\sigma}=i\sigma\ket{yz,\sigma}$ and $L_zS_z\ket{yz,\sigma}=-i\sigma\ket{xz,\sigma}.$ The $S_z$ operator is even under spatial inversion but $L_z$ changes sign. As the two graphene valleys are related by inversion, the effective SOC interaction has opposite sign for $\mbfK$ and $\mbfK'$ resulting in a $\tau_z$ dependence.

\begin{figure}
    \centering
    \includegraphics[width=0.7\linewidth]{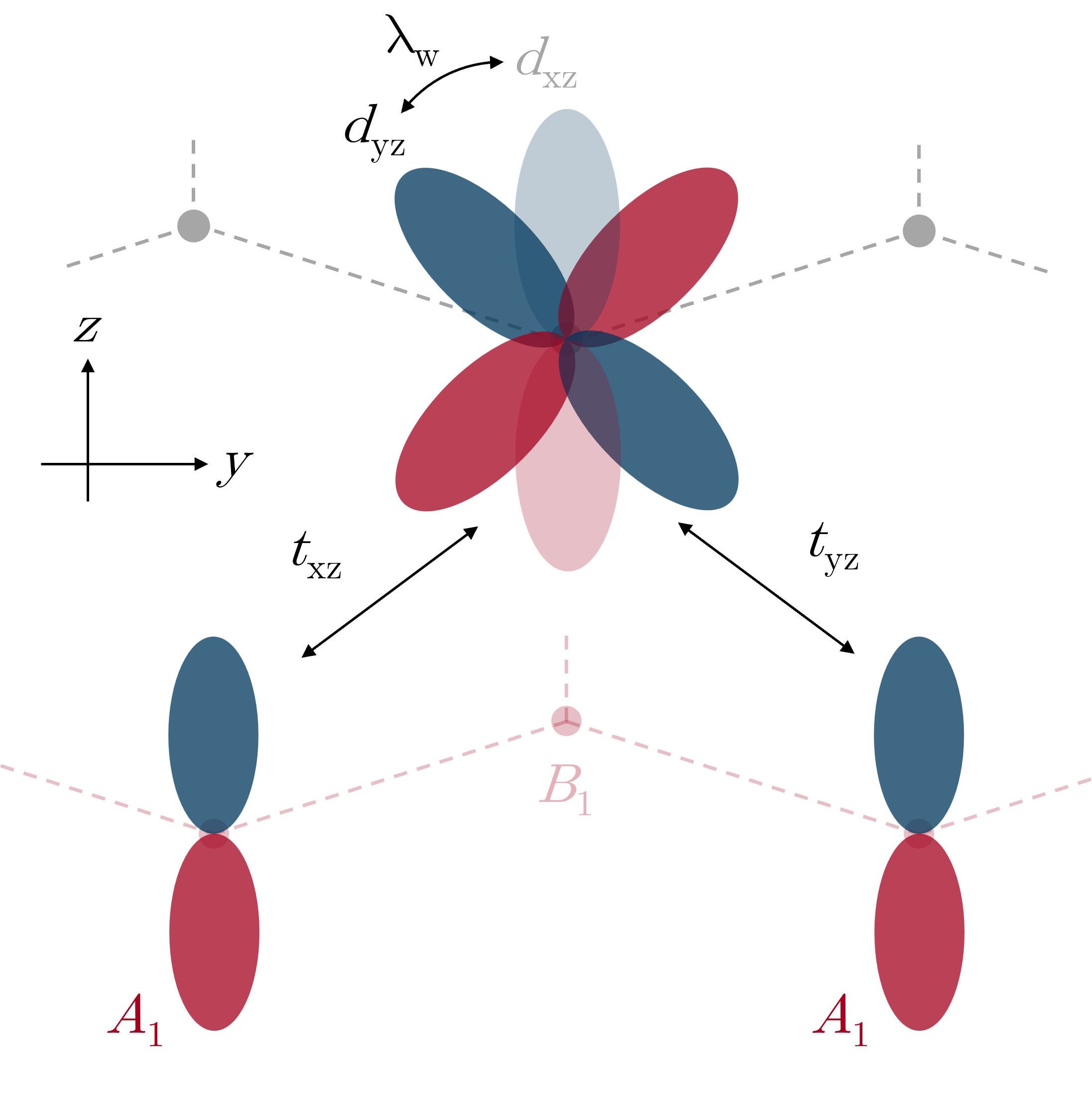}
    \caption{Schematic depiction of the ${A_1\rightarrow d_{xz/yz}\rightarrow A_1}$ (graphene $\rightarrow$ tungsten$ \rightarrow$ graphene) hopping processes. The WSe$_2$ is placed above the BBG, proximate to layer $l=1.$ Electrons from the graphene $A_1$ $p_z$-orbitals hop between themselves via an intermediate hopping to the tungsten $d_{xz/yz}$ orbitals, who are mixed by the atomic SOC $\lambda_w.$ Red and blue colouring of the orbitals correspond to the sign of the wavefunction.}
    \label{ProximityFig}
\end{figure}

To see how this comes about, we consider a simplified model of the BBG-WSe\textsubscript{2} heterostructure. We take as our starting point the triangular sublattice of the $A_1$ sites. Focusing on momentum ${{\bf K} = (0, K_y)},$ we consider a hopping process involving the $p_z$-orbitals of $A_1$ and the $d$-orbitals of tungsten along the $y$-axis (see Fig. \ref{ProximityFig}).

Imagine an $A_1$ electron hops to a $d_{xz}$ orbital, whereupon it is mixed with $d_{yz}$ by the atomic SOC $\lambda_w,$ and then hops back to $A_1.$ Such process leads to  the effective SOC term
\begin{equation}\label{effectiveSOC}
    \mathcal{H}_\lambda\sim\frac{ t_{xz} \lambda_w t_{yz}}{E_{pd}^2} \sigma \delta_{\sigma\sigma'},
\end{equation}
where the standard perturbation theory has been employed assuming that $E_{pd}$, the atomic energy difference between $p_z$ and $d_{xz}/d_{yz}$ is much larger than $\lambda_w$, $t_{xz},$ and $t_{yz}$. The $t_{yz}$ hopping changes sign under $y\rightarrow-y,$ but $t_{xz}$ does not (see Fig. \ref{ProximityFig}). Thus, from Eq. (\ref{effectiveSOC}) it becomes clear that the SOC interaction takes opposite sign in opposite valleys (i.e. under $K_y \rightarrow -K_y$). Note that the tungsten atoms must be displaced slightly along the $x$-direction to allow for finite $t_{xz},$ which would otherwise be zero by symmetry.

Introducing $\Psi_{\bf p} = (c_{+, \mbfp,\uparrow},c_{-,\mbfp,\uparrow}, c_{+,\mbfp,\downarrow},c_{-,\mbfp,\downarrow})^T,$ the effective Ising SOC Hamiltonian acting in the $A_1$ subspace reduces to
\begin{equation}
    \mathcal{H}_\lambda=\lambda \sum_\mbfp \Psi_\mbfp^\dagger (\sigma_z \tau_z) \Psi_\mbfp,
\end{equation}
where ${\vec \sigma}$ is another Pauli matrix acting in the spin degree of freedom and $\lambda \sim t_{xz} \lambda_w t_{yz} /{E_{pd}^2}$. Estimates for the magnitude of $\lambda$ vary, but generally are on the order of $1-10$ meV \cite{David2019,Gmitra2015PRB,WangZ2015,WangZ2016,WangD2019,Szentpeteri2024,SOCvsTheta}. Within our patch model, the SOC strength is approximated as constant in magnitude, with equal and opposite values in the $\mbfK$ and $\mbfK'$ valleys. The constant-magnitude approximation is reasonable, as the Fermi surface pockets are small. For the same reason, we neglect Rashba SOC. The effective low-energy single-particle Hamiltonian is then given by ${\mathcal{H}_0 + \mathcal{H}_\lambda}$.

Our simplified treatment does not take into account the lattice geometry of the WSe$_2$/BBG heterostructure. In particular, due to the lattice mismatch between the graphene and the WSe$_2,$ the hopping process described above may only be possible in certain domains of the sample. However, even with a relative twist angle between the lattices, Ising SOC is expected to be generic; only the magnitude of $\lambda$ is affected by twist angle \cite{David2019}. A more sophisticated study taking into account the atomic sites of the supercell using first-principle methods reported an effective Ising SOC near the $K$ points consistent with the above result \cite{Gmitra2015PRB,SOCvsTheta}. Another microscopic model for the origin of Ising SOC is proposed in Ref. \cite{ChouYZ2022}.

\subsection{Fermi surface topology}
\begin{figure}
    \centering
    \includegraphics[width=\linewidth]{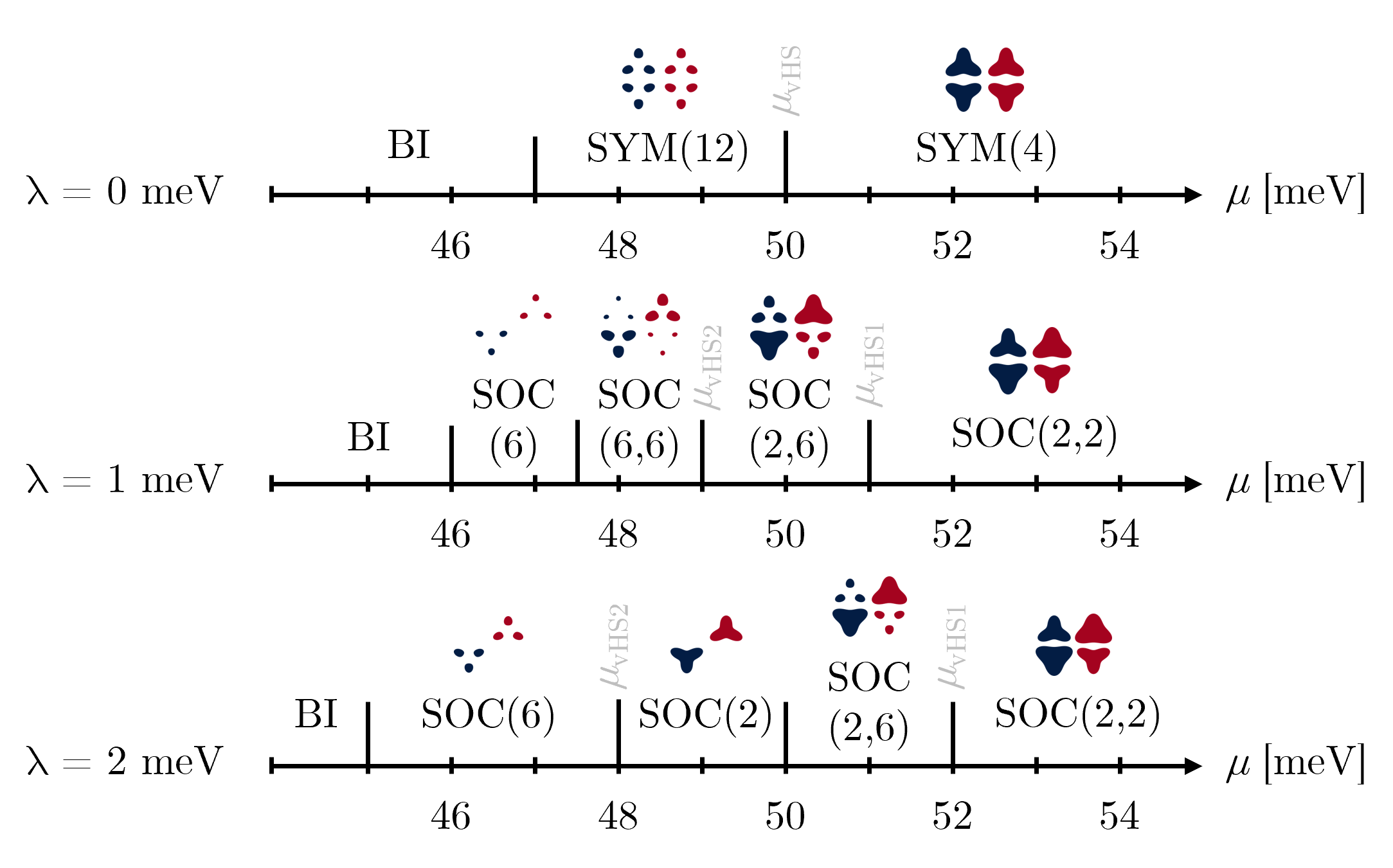}
    \caption{Evolution of Fermi surface (FS) topology for the effective tight binding model ${\cal H}_0 + {\cal H}_\lambda$ as a function of $\mu$ for three different values of $\lambda.$ In each schematic, the top and bottom rows correspond to $K$ and $K',$ respectively, while colour denotes spin. SYM and SOC refers to the symmetric band without SOC and to the band with SOC, respectively. The index inside the bracket counts the number of FS pockets. When there are two different sized pockets, two indices are used for large and small pockets. BI stands band insulator. We set $D=$ 50 meV. The locations of the van Hove singularities are also indicated.}
    \label{FS}
\end{figure}

It is useful to establish how the Fermi surface (FS) changes in the presence of SOC. In Fig. \ref{FS}, we present the FS topology for the tight-binding model as a function of the chemical potential $\mu$, for three choices of the SOC strength $\lambda.$ We focus on electron-doping into the $E^{(3)}$ band. Our results apply equally for hole-doping into the $E^{(2)}$ band due to particle-hole symmetry.

In the absence of SOC, the system possesses four degenerate bands (spin $\times$ valley). There is a vHS at $\mu_{\rm vHS}\approx D$ (here we take $D = 50$ meV). As the chemical potential moves through the vHS, each of the four degenerate Fermi surfaces separate into three Fermi pockets. As $\mu$ increases, the FS topology changes from $\rm{SYM}(12)$ to $\rm{SYM}(4).$ Here, `SYM' indicates a FS topology without any broken symmetry, and the index counts the number of equal-sized FS pockets. For chemical potentials below the band edge, the system is a band insulator (BI).

When $\lambda\neq 0,$ the four-fold degeneracy is lowered to two sets of two-fold degenerate bands. The two-fold degenerate vHSs occur at $\mu_{\rm vHS1} \equiv D + \lambda$ and $\mu_{\rm vHS2} \equiv D - \lambda.$ The SOC thus allows for various other FS topologies. In Fig. \ref{FS}, `SOC' refers to the bands with finite $\lambda$ and the two indices $(\alpha,\beta)$ count the number of unequal-sized FS pockets in order of decreasing size. For example, $\rm{SOC}(2,6)$ labels the topology consisting of 2 larger and 6 smaller Fermi pockets. Some topologies such as $\rm{SOC}(6,6)$ and $\rm{SOC}(2)$ are only available for $2\lambda<W$ and $2\lambda>W,$ respectively, where $W\approx 3.4$ meV is the bandwidth between the band edge and the local maximum. An extended look at these FS topologies is given in Appendix \ref{smallSOC}.

\section{Low-energy effective interaction Hamiltonian}\label{Interacting}

Now that we have established the low-energy tight-binding model ${\cal H}_0+{\cal H}_{\lambda}$, including the Ising SOC, we turn to the low-energy interactions of doped electrons near the valleys to capture the possible broken-symmetry states. The interacting Hamiltonian for electrons on the non-dimer $A_1$ and $B_2$ sites, including on-site, NN, and NNN Hubbard interactions, is given by
\begin{eqnarray}
 \mathcal{H}_{\rm int} &= &  U_{A_1}\sum_{i \in A_1} {n}_{i,\uparrow} {n}_{i, \downarrow} + U_{B_2}\sum_{i \in B_2} { n}_{i,\uparrow} {n}_{i\downarrow} \nonumber\\
 & &  + \; V_0\sum_{\langle i,j\rangle} { n}_i { n}_j
 + V_0^\prime
 \sum_{\langle\langle i,j\rangle\rangle} { n}_i { n}_j,    
\end{eqnarray}
where ${n}_{i \in A_1} =\sum_\sigma {a}^\dagger_{1,i,\sigma} {a}_{1,i,\sigma}$ is the electron density operator for $A_1$ sites, and similarly ${n}_{j \in B_2} =\sum_\sigma {b}^\dagger_{2,j,\sigma} {b}_{2,j,\sigma}$ for $B_2$ sites. Here, $U_{A_1}$ and $U_{B_2}$ are the on-site Hubbard interactions at $A_1$ and $B_2,$ respectively. $V_0$ is the NN density-density interaction between electrons on $A_1$ and $B_2$ which form the emergent honeycomb lattice. $V_0^\prime$ is the NNN interaction between $A_1$ and $A_1$ or $B_2$ and $B_2.$ All interactions are assumed repulsive. We note that this model is also applicable for hole doping into the valence band, as demonstrated in Appendix \ref{PHT}.

At half-filling, the low-energy triangular $B_2$ sublattice is fully occupied due to the displacement field $D$. When the system is lightly electron-doped, the additional electrons are constrained to the high-energy $A_1$ sublattice. The effective interacting Hamiltonian for $A_1$ electrons can be derived using a Schrieffer-Wolff transformation, similar to the procedure pioneered in Ref. \cite{spintriplet} for a honeycomb lattice with a staggered sublattice potential.

Including occupations of up to three fermions per upper triangle, the resulting low-energy interacting Hamiltonian is found to be
\begin{align*}
    \mathcal{H}^{\rm{eff}}_{\rm int} & = \sum_{\expect{ij}, \sigma}\left[\frac{\Tilde{t}}{2}\left(n_i +n_j\right)\right]\left(a_{1,i,\sigma}^\dagger a_{1,j,\sigma} + h.c. \right) \\ & +\gamma \sum_{ijk\in \triangle,\sigma} \left(a_{1,i,\sigma}^\dagger n_k a_{1,j,\sigma} +P_{ijk}\right) \\ &  + U\sum_i n_{i\uparrow} n_{i\downarrow} + V\sum_{\expect{ij}} n_{i} n_{j}, \numberthis
\end{align*}
where ${\bar D}= 2D - U_{A_1}+3 V_0 - 12 V_0^\prime$, $\gamma = t_0^2/({\bar D} + 2 V_0^\prime) - t_0^2/({\bar D}+V_0) \; >0,$ and $\Tilde{t} = t_0^2/({\bar D}+2 V_0^\prime) + t_0^2/({\bar D}+U_{A_1} + 3 V_0^\prime) - 2t_0^2/({\bar D}+V_0) \; > 0.$ Here, $t_0$ is the effective bandwidth of band $E_{\bf k}^{(3)}$ near the valley. The renormalized interactions go as $U \sim U_{A_1}$ and $V \sim V_0^\prime$, assuming that $D \gg U_{A_1} >  V_0 > 2 V_0^\prime > t_0$. The full expressions for the renormalized $U$ and $V$ are shown in Appendix \ref{renorm}. The interaction $\gamma \rightarrow 0$ as $V_0, V_0^\prime \rightarrow 0,$ and corresponds to correlated hopping between three sites of $i,j,k \in A_1$ due to the presence of occupied $B_2$ sites in the middle of a triangle formed by NN $A_1$ atoms ($P_{ijk}$ refers to permutation among the $ijk$ forming the triangle). Note that we have not included the Ising SOC $\lambda$ in this procedure as $\lambda \ll (D, t_0)$.

Focusing on momenta near the valleys $\tau \in \{ +, -\}$, and retaining fermionic modes $a_{1, \tau K +\mbfp,\sigma} \equiv c_{\tau,\mbfp,\sigma} $, a Fourier transform of the above model results in
\begin{equation}
    \mathcal{H}_{\rm int}^{\rm eff} = \frac{1}{N}\sum_{\mbfq} g_1\,\rho_{+}(\mbfq) \rho_{-}(-\mbfq) + g_2\,\mathbf{s}_+(\mbfq)\cdot\mathbf{s}_{-}(-\mbfq),
\end{equation}
where
\begin{equation}
    \rho_\tau(\mbfq) \equiv \sum_{\mbfp,\sigma} c^\dagger_{\tau, \mbfp, \sigma} c_{\tau,\mbfp+\mbfq,\sigma}
\end{equation}
and
\begin{equation}
    \mathbf{s}_\tau(\mbfq) \equiv \frac{1}{2}\sum_{\sigma\sigma^\prime}\sum_{\mbfp}  c_{\tau,\mbfp,\sigma}^\dagger\boldsymbol{\sigma}_{\sigma\sigma'}c_{\tau,\mbfp+\mbfq,\sigma'}.
\end{equation}
The parameters $g_1$ and $g_2$ are given by
\begin{eqnarray} \label{g1g2_orig}
g_1 &=& \frac{1}{2}(U + 15V - 24 \gamma - 6 {\Tilde t}) \;  > 0,\nonumber\\
g_2 &=& -2 (U - 3V + 12 \gamma - 6 {\Tilde t }) \;  < 0.
\end{eqnarray}
As we will discuss in Sec. \ref{disco}, the WSe$_2$ substrate may modify the strengths of $g_1$ and $g_2$ in addition to inducing the Ising SOC. Below we consider a full Hamiltonian ${\cal H}_0 + {\cal H}_{\lambda} + {\cal H}_{\rm int}$ and explore possible broken-symmetry states within MF theory. We limit our study to broken-symmetry states with zero centre-of-mass momentum ${\bf q}=0$.

\section{Mean-field order parameters}\label{MFOP}

Note that $g_1$ is repulsive while $g_2$ is attractive, indicating that they act like the density-density and Hund's coupling between the valleys. Let us examine the attractive channels. Since the interactions can be reformulated as $- g_1 (\rho_+ - \rho_-)^2$ and $g_2 (\mathbf{s}_+ +\mathbf{s}_-)^2,$ with $g_1>0$ and $g_2 <0$, they lead to attractive channels for valley polarization ($P_z$) and spin polarization (${\bf M}$) with ${\bf q}=0$:
\begin{equation}
     P_z  \equiv \frac{1}{2} \langle \left(\rho_+ - \rho_-\right) \rangle, \;\;\; 
    \mathbf{M}  \equiv \langle \left( \mathbf{s}_+ +\mathbf{s}_- \right) \rangle.
\end{equation}
Intervalley-order parameters $P_x$ and $P_y$ are also possible. We collect the three isospin components into a vector
\begin{equation}
    \mathbf{P} = \frac{1}{2}\sum_{\tau\tau',{\bf p},\sigma} \langle \left( c^\dagger_{\tau,{\bf p}, \sigma} \left(\boldsymbol{\tau}\right)_{\tau\tau'} c_{\tau',{\bf p},\sigma}\right) \rangle.
\end{equation}
Given the trigonal warping associated with the $\varepsilon^a_{\bf p}\,\tau_z$ term in Eq. (\ref{KineticTau}), we expect $P_z$ to be different from $P_x$ and $P_y.$ Spin and valley polarization are treated on equal footing, in that spin polarization is valley-independent and valley polarization is spin-independent.

We also consider the particle-particle channel. The MF interaction strength for the spin-singlet is $(g_1 - \frac{3}{4} g_2)/2> 0$, i.e. repulsive due to $g_1>0$ and $g_2 < 0.$ For this reason we do not expect the spin-singlet to form. However, the spin-triplet has an attractive interaction \cite{spintriplet} and is associated with the order parameter
\begin{equation}
    \boldsymbol{\Delta} = \frac{1}{2}\sum_{\tau\tau',\mbfp,\sigma\sigma'} \langle c_{\tau,\mbfp,\sigma}\left( i\sigma_y\boldsymbol{\sigma} \cdot {\hat d} \right)_{\sigma\sigma'} \left(i {\tau_y}\right)_{\tau\tau'}  c_{\tau',-\mbfp,\sigma'} \rangle.
\end{equation}
Here, ${\hat d}$ denotes the $d$-vector, which is perpendicular to the spin of the Cooper-pair condensate, i.e. ${\hat d} \cdot {\bf S}|\psi_{sc}\rangle = 0,$ where ${\bf S}$ is the total spin of the Cooper pair and $|\psi_{sc}\rangle$ is a spin-triplet SC state. For example, when ${\hat d} = {\hat z}$, the triplet occurs in $S_z=0$. Note that the above spin triplet is a valley-singlet SC, and thus the antisymmetric wave function condition is satisfied under the exchange of the two valleys. 

The interacting MF Hamiltonian can be recast as
\begin{align*}\label{recasttt}
    \mathcal{H}_\text{int}^{\rm MF} & = V_P \sum_{\tau\tau',\bf p,\sigma} c^\dagger_{\tau,{\bf p},\sigma} \left(\boldsymbol{\tau}\cdot {\bf P} \right)_{\tau\tau'}c_{\tau',{\bf p},\sigma}  \numberthis  \\
    + V_{M} & \sum_{\tau,\bf p,\sigma\sigma'}  c_{\tau,\mbfp,\sigma}^\dagger \left( \boldsymbol{\sigma} \cdot {\bf M} \right)_{\sigma\sigma'}c_{\tau,\mbfp,\sigma'} \\
    +\frac{V_T}{2} &\, \sum_{\tau\tau',\mbfp,\sigma\sigma'} c^\dagger_{\tau,\mbfp,\sigma} \left( {\vec \Delta} \cdot \boldsymbol{\sigma} i\sigma_y \right)_{\sigma\sigma'} \left(i {\tau_y}\right)_{\tau\tau'}  c^\dagger_{\tau',-\mbfp,\sigma'},
\end{align*}
where the MF interaction strengths are given by
\begin{align*}\label{pol_int}
     V_{P} & = -g_1, \\
     V_{M} & = \frac{1}{4}g_2,\\
    V_{T} & = \frac{1}{2}\left(g_1 + \frac{1}{4} g_2\right). \numberthis 
\end{align*}
The attractive interaction is obtained for the spin triplet when ${-g_2>4g_1}.$ Rewriting $g_1$ and $g_2$ in terms of ${\tilde t}$, $\gamma$, $U$, and $V$ [Eq. (\ref{g1g2_orig})], we have $2V_T = 9V - 18 \gamma,$ indicating that the attractive interaction for the SC originates from $\gamma$, the assisted hopping interaction. Meanwhile, for the spin- and valley-polarizing channels, we obtain attractive interactions for any $g_1 >0$ and $g_2 < 0$. The representation of the full MF Hamiltonian in terms of spin, valley, and particle-hole degrees of freedom is given in Appendix \ref{NambuAppendix}. Solving for the order parameters, ${\bf P,}$ ${\bf M},$ and $\boldsymbol{\Delta}$ self-consistently at zero temperature, we obtain the MF results presented in the next section.

\section{Mean-field results and implications}\label{MF}

\subsection{Phase diagrams}

\begin{figure*}
    \centering
    \includegraphics[width=\linewidth]{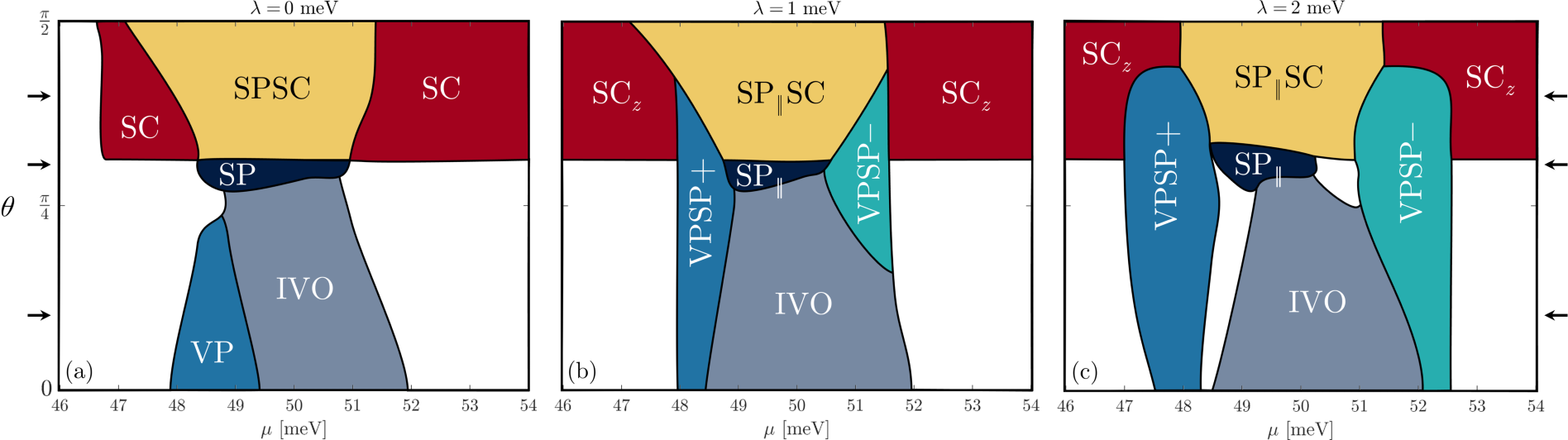}
    \caption{Phase diagrams as a function of $\mu$ and $\theta$ for (a) $\lambda=0,$ (b) $\lambda=1$ meV, and (c) $\lambda=2$ meV. The interaction strengths are parameterized as $g_1=g \cos(\theta)$ and $g_2/4= - g \sin(\theta)$, so that $\theta=\pi/4$ corresponds to $|V_P/V_M|=1$ and $V_T$ becomes attractive. We set $g= 7$meV. The arrows indicate the location of the linecuts in Fig. \ref{MFplots}. Disordered phases are shown in white.}
    \label{Thetaplot}
\end{figure*}

The phase diagrams are presented in Fig. \ref{Thetaplot} for three choices of the Ising SOC $\lambda.$ The results are given as a function $\mu$ and $\theta,$ where $\theta$ parameterizes the interactions according to $g_1 = g \cos(\theta),$ $g_2/4=-g\sin(\theta).$ As is evident from Eqs. (\ref{recasttt}) and (\ref{pol_int}), $g_1$ favours valley polarization (VP) and intervalley order (IVO), characterized by finite $P_z$ and $(P_x,P_y),$ respectively. The attractive $g_2$ favours spin polarization (SP) and spin-triplet superconductivity (SC). With this parameterization, $\theta\rightarrow 0$ corresponds to $|g_2/g_1|\ll 1,$ where ordering in the valley degree of freedom is anticipated. $\theta\rightarrow\pi/2$ corresponds to $|g_2/g_1|\gg 1,$ where superconductivity and ordering in the spin degree of freedom is expected. At $\theta=\pi/4,$ the VP and SP interactions become equal ($V_P/V_M=1$) and the SC interaction $V_T$ becomes attractive. 

We set $g=7$ meV throughout, which is the minimum interaction strength that allows for VP and SP to develop. This is well within the range appropriate for a mean-field theory, where the scale of interactions should be much smaller than the separation to the nearest bands. The present theory requires $g$ much smaller than $2D$ and $t_\perp$ (see Fig. \ref{BZ}), which are on the order of $\sim 100$ meV. If $g$ is increased, interaction effects will begin to dominate the kinetic terms in the Hamiltonian and the ordered phase space will be enlarged. However, the competition between the various observed phases remains intact and our conclusions are unchanged. Our results are also robust to the particular choice of hopping amplitudes. We have performed calculations using the parameters reported in Jung and MacDonald \cite{JungAndMacDonald2014} and compared them with Fig. \ref{Thetaplot}, finding similar outcomes. Numerical details of our MF calculations can be found in Appendix \ref{numericaldeets}. 

We begin with a high-level overview of the results before elaborating on the details. Consider first Fig. \ref{Thetaplot}(a), where $\lambda=0$. For small $\theta,$ IVO and VP occur near the vHS at $\mu_{\rm vHS}\approx 50$ meV. As $\theta$ increases, IVO transitions into SP, followed by a coexistence of SP and SC (SPSC). The transition from IVO to SP does not occur precisely at $\theta=\pi/4$ because IVO can gap out the FS while SP cannot. Away from $\mu_{\rm vHS},$ the spin-triplet SC occurs for $\theta>\pi/4.$ The appearance of a sharp onset of the SC state as $\theta$ increases is unphysical; the SC order parameter will develop gradually beyond $\theta = \pi/4,$ except near the top-left corner of (a) where the chemical potential passes below the band edge. For $\theta<\pi/4,$ the system remains in the unpolarized state away from the vHS.

Fig. \ref{Thetaplot}(b) gives the result for $\lambda = 1$ meV. As previously noted, the Ising SOC splits the original vHS at $\mu_{\rm vHS}$ into two distinct values at $\mu_{\rm vHS} \pm \lambda.$ Here we have $\mu_{\rm vHS_1}=51$ meV and $\mu_{\rm vHS_2}=49$ meV. The phases denoted VPSP$\pm$ (coexisting valley and spin polarization) develop about $\mu_{\rm vHS_1}$ and $\mu_{\rm vHS_2}.$ IVO persists in the intermediate region. Interestingly, with finite $\lambda$ the VPSP phase begins to encroach upon the phase space of SC and SPSC. A reduction in the occupied phase space of SC near the vHS due to SOC has been reported in Ref. \cite{ZhangY2024}. Another effect of the Ising SOC is to pin the orientation of the $d$-vector and the magnetization axis, which we indicate in Fig. \ref{Thetaplot} by the subscripts $\rm{SC}_z$ and $\rm{SP}_\|$ (see below). The SC regions occurring away from the vHSs are not enhanced by the SOC, except in the top-left corner. However, this is simply a result of the splitting induced by $\lambda,$ pushing off one set of bands towards lower energies. 

\begin{figure*}
    \centering
    \includegraphics[width=0.9\linewidth]{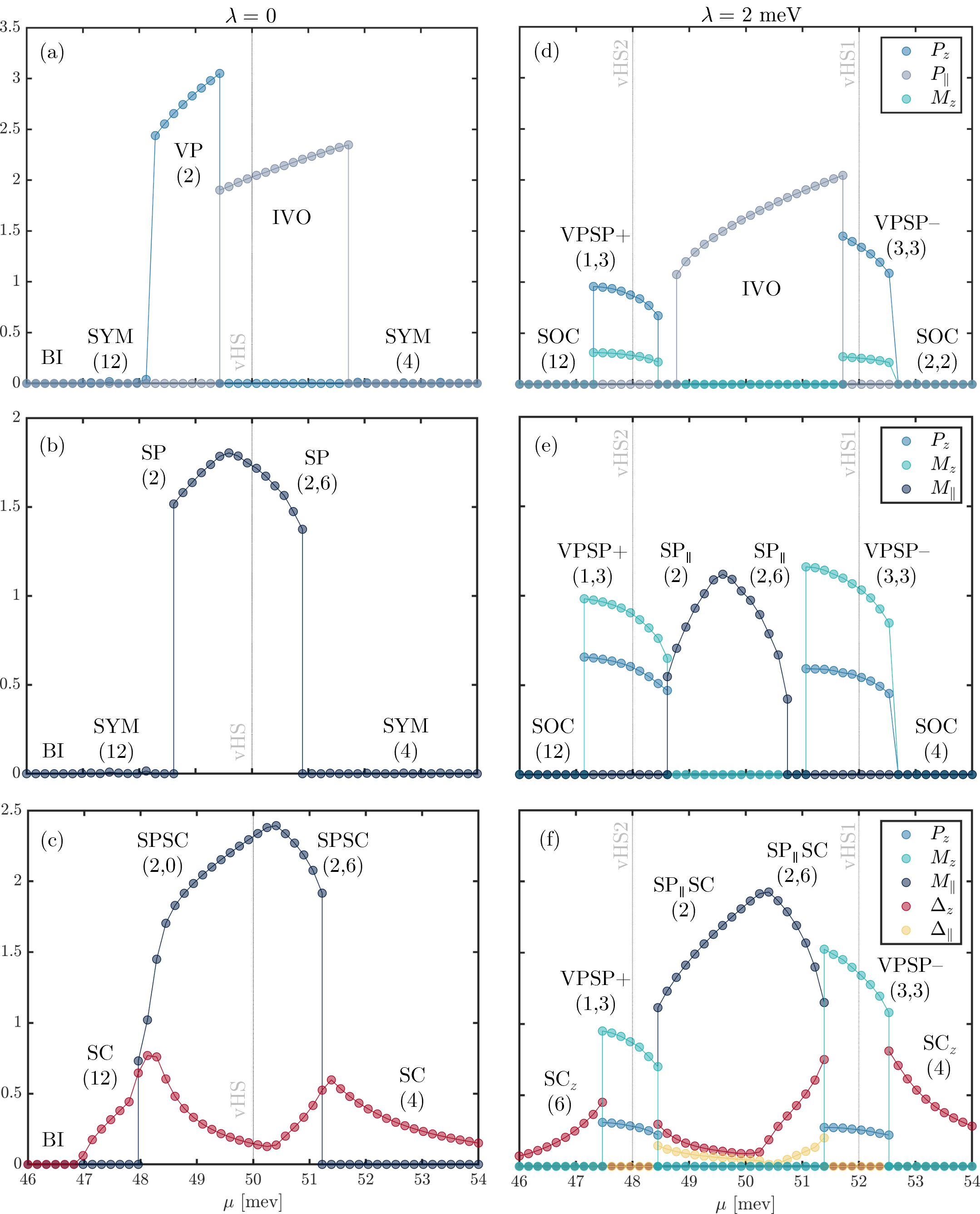}
    \caption{Order parameter amplitudes (in unit of meV) as a function of $\mu$ for three representative linecuts of Fig. \ref{Thetaplot}. From top to bottom, $\theta$ increases. Left and right column are for $\lambda =0$ and $\lambda =2$ meV, respectively. IVO, VP, SP, VPSP, SC, and SPSC refer to intervalley-ordered, valley-polarized, spin-polarized, valley- and spin-polarized, superconducting, and spin-polarized superconducting states, respectively. Top row: $\theta = \pi/10,$ leading to $V_P=-6.7$ meV, $V_M=-2.2$ meV, and $V_T > 0$. Middle row $\theta = 5\pi/16,$ leading to $V_P=-3.9$ meV and $V_M=-5.8$ meV. Bottom row: $\theta = 2\pi/5,$ leading to $V_P = -2.2$ meV, $V_M = -6.7$ meV, and $V_T = -2.2 $ meV. The triplet amplitude has been multiplied by a factor of 2 for visibility. For the $\lambda=0$ cases (a)--(c), $M_x=M_y\equiv M_\| = M_z$ and only $M_\|$ is shown for clarity. Similarly, $P_x=P_y\equiv P_\|$ for IVO. The location of the vHSs are indicated.}
    \label{MFplots}
\end{figure*}

As $\lambda$ is increased to 2 meV in Fig. \ref{MFplots}(c), the VPSP phases encroach further upon SC and SPSC. Pockets of disorder emerge at the phase boundaries between IVO and VPSP. This is because $\mu_{\rm vHS_1}$ and $\mu_{\rm vHS_2}$ (now at 52 meV and 48 meV, respectively) are driven further apart as $\lambda$ is increased, leaving intermediate densities at very low DOS. Pure spin polarization SP$_{\parallel}$ is confined to the phase space near $\theta=\pi/4,$ between $\mu_{\rm vHS_1}$ and $\mu_{\rm vHS_2}.$

We now examine these results more carefully, by considering linecuts of the phase diagrams for three representative $\theta$ indicated by the black arrows in Fig. \ref{Thetaplot}. They are presented in Fig. \ref{MFplots}, in which the three rows correspond to the three linecuts. The left column (a)--(c) shows the results for $\lambda=0$ and the right column (e)--(f) shows the results for $\lambda=2$ meV.  We will consider these two cases in turn to explore the effects of Ising SOC.

\subsection{Effects of interactions without Ising SOC}\label{woSOC}

For $\lambda=0,$ the lone vHS is located at $\mu_{\rm vHS} \approx 50$ meV. In Fig. \ref{MFplots}(a), the linecut corresponds to $\theta=\pi/10,$ for which $g_1$ is the dominant interaction. In this regime, VP and IVO occur due to $V_P=-6.7$ meV dominating over $V_M=-2.2$ meV, with $V_T$ being repulsive. The abrupt onset of the VP order parameter from the symmetric state SYM(12) as $\mu$ approaches $\mu_{\rm vHS}$ indicates a first-order transition resembling the Pomeranchuk (or nematic) instability. The density difference between valleys leads to a splitting of the vHS into two distinct vHSs (one below and one above $\mu_{\rm vHS}$), thereby lowering the system’s energy \cite{Puetter2007,Khavkine2004}. There is then a first-order transition to IVO, developing when the chemical potential is tuned higher into the band. A similar transition has been observed in rhomboherdal trilayer graphene \cite{Arp2023}.

In the IVO state, the $P_x$ and $P_y$ solutions are degenerate and are collectively denoted by $P_\|.$ IVO hybridizes the $K$ and $K'$ bands, while VP can only shift them up and down. Hybridization can be energetically favourable if it gaps out the Fermi surface. As discussed above, the $\rm{SU}(2)$ isospin symmetry of $\mathbf{P}$ is broken by the $\tau_z$-dependence in the electronic dispersion, and the degree of asymmetry is parameterized by the trigonal warping $t_3$. At $t_3=0,$ the $\rm{SU}(2)$ symmetry is restored and we recover the degenerate solution $P_x=P_y=P_z.$ For finite $t_3,$ we have a first-order transition between VP and IVO. The critical $\mu$ depends on details of the calculation, namely the strengths of $t_3$ and $V_P.$

The linecut in Fig. \ref{MFplots}(b) corresponds to $\theta=5\pi/16,$ where $V_M=-5.8$ meV dominates over $V_P=-3.9$ meV. In comparison with \ref{MFplots}(a), the VP order parameter has been replaced by SP. Like VP, the SP order parameter develops around $\mu_{\rm vHS}$. The first-order transition from SYM(12) or SYM(4) highlights the significance of trigonal warping near the vHS, which differs from the conventional second-order Stoner instability. Since $\lambda=0,$ the three magnetization components $(M_x,M_y,M_z)$ are equivalent and yield degenerate solutions. $V_T$ is also attractive in this regime, but SC is absent as the interaction is too weak $(V_T \approx -1$ meV).

In Fig. \ref{MFplots}(c), $\theta=2\pi/5$ again leaves $V_M$ as the dominant interaction. Compared with \ref{MFplots}(b), the SP phase is enlarged due to larger $V_M=-6.7$ meV ($V_P=-2.2$ meV). Moreover, $V_T=-2.2$ meV is now appreciably attractive and SC develops over a wide range of $\mu.$ In the pure SC regime away from the vHS, the orientation of the $d$-vector is arbitrary. Close to the vHS, the SC order parameter coexists with SP (hence SPSC). Within the coexistence region, the $d$-vector is pinned to the plane perpendicular to the (spontaneously chosen) magnetization axis. The SC order parameter develops continuously from the unpolarized state, however, when finite SP occupies the phase space near $\mu_{\rm vHS}$, the strength of the SC order parameter is suppressed. One might have expected the large DOS available at the vHS to enhance superconductivity, however the spontaneous spin splitting pushes the vHS away from the pairing surface, resulting in a suppression of the superconducting gap.

The modification of the FS topologies in the ordered phases are labelled in Fig. \ref{MFplots} according to the scheme described in Sec. \ref{Theory}. Some examples are shown in Appendix \ref{evobandtop}. We acknowledge that order-parameter amplitudes are typically overestimated by MF theory, and so our labelling of the FS topologies in the symmetry-broken phases may differ beyond MF.

\subsection{Effects of Ising SOC}\label{wSOC}

Let us now turn to the results of Fig. \ref{MFplots}(e)--(f) with nonzero $\lambda=2$ meV. The locations of the shifted vHSs in the finite $\lambda$ case ($\mu_{\rm vHS_1}$ and $\mu_{\rm vHS_2}$) are indicated by a pair of vertical lines. Due to the symmetry of the Ising SOC, $M_x$ and $M_y$ ($\Delta_x$ and $\Delta_y$) are degenerate solutions, which we represent collectively by $M_{\parallel}$ (${\Delta}_\parallel$).
 
In Fig. \ref{MFplots}(d), $V_P$ is dominant over $V_M$ and $V_T$. Compared with the $\lambda=0$ case of \ref{MFplots}(a) at the same $\theta,$ the VP state now occupies the phase space near $\mu_{\rm vHS_1}$ and $\mu_{\rm vHS_2}$. This behaviour suggests that the finite-VP state with $\lambda\neq 0$ closely resembles the Pomeranchuk instability associated with the $\lambda=0$ case. Additionally, we now observe a finite spin polarization along the $z$-axis within the VP phase (hence VPSP) induced by the Ising SOC. This is due to the linear coupling between $P_z$ and $M_z$ through the term $\lambda \tau_z \sigma_z$. Since $\lambda > 0$, naively one expects opposite signs for $P_z$ and $M_z.$ This expectation is met near $\mu_{\rm vHS_1}$ (denoted VPSP$-$), but near $\mu_{\rm vHS_2}$ the same sign of $P_z$ and $M_z$ occurs (denoted VPSP$+$). The explanation has to do with which set of SOC-polarized bands participate in the Pomeranchuk transition (see Appendix \ref{evobandtop}). The IVO is less sensitive to the shifted vHSs, and persists in the intermediate region between $\mu_{\rm vHS_1}$ and $\mu_{\rm vHS_2}$. In principle, the IVO phase can support superconductivity, but in the regime where $V_T$ is attractive, $V_P$ is small and IVO does not develop.

In Fig. \ref{MFplots}(e), the strong tendency to develop VP near the vHSs persists, even though $V_M$ is dominant over $V_P$. Similar to case \ref{MFplots}(d), VPSP$\pm$ emerges near $\mu_{\rm vHS_1}$ and $\mu_{\rm vHS_2}$, although now the SP amplitude is larger than VP. Here, VP is induced by $M_z$ through the combined effects of Ising SOC and proximity to the vHS. Away from the vHSs, the SP state persists without the presence of VP, but only $M_x$ and $M_y$ are finite (denoted SP$_{\parallel}$). The in-plane and out-of-plane components of the magnetization are inequivalent because of the Ising SOC, which acts along the $z$-axis. In the $\rm{SP}_\|$ phase, the valley-resolved spin projections are canted, with equal and opposite $M_z$ components for $K$ and $K'.$ The in-plane components are equal in $K$ and $K'$, resulting in a net in-plane magnetization for the system. The canting angle depends on the relative magnitudes of $\lambda$ and $M_\|.$ This phase has been observed in the self-consistent Hartree-Fock calculations of Ref. \cite{hartreefock1}.

The reemergence of VP near the vHSs is also evident in Fig. \ref{MFplots}(f). VPSP appears near $\mu_{\rm vHS_1}$ and $\mu_{\rm vHS_2},$ overtaking SPSC when compared to the $\lambda=0$ case \ref{MFplots}(c) at the same $\theta.$ The presence of VP, which differentiates the FS in the two valleys, strongly suppresses spin-triplet SC. Our analysis indicates that transitions between VPSP and SC/SPSC are first-order, with no coexistence of VP and SC. When VPSP subsides away from the vHSs, the SC recovers. When $\mu$ is located between $\mu_{\rm vHS_1}$ and $\mu_{\rm vHS_2},$ the Ising SOC results in the SP$_{\parallel}$SC phase, with inequivalent $\Delta_\parallel$ and $\Delta_z$. The Ising SOC favours the $\Delta_z$ component because $\Delta_z$ pairs opposite spins between opposite valleys with an equal-sized partner FS. The $\Delta_{\parallel}$ component therefore weakens, but is still finite within SP$_{\parallel}$SC. The pure SC denoted by $\rm{SC}_z(4)$ develops for $\mu > \mu_{\rm vHS_1}$ where only $\Delta_z$ is finite, and where (4) refers to the number of underlying FS pockets before the SC gap develops. A second region of spin-unpolarized superconductivity $\rm{SC}_z(6)$ develops when $\mu < \mu_{\rm vHS_2},$ close the band edge of the SOC-split bands. The underlying FS consists of 6 Fermi pockets with opposite spin and opposite valley quantum numbers. The SC gap forming near the Fermi level in $\rm{SC}_z(4)$ and $\rm{SC}_z(6)$ is depicted in Appendix \ref{evobandtop}.

These results suggest that the SC phase is not significantly enhanced by the Ising SOC. Instead, SOC favours VPSP close to a vHS, while SC emerges only away from a vHS. The VPSP phase segments the superconducting regime into several disconnected regions, which may help to explain the two distinct SC domes reported in Ref. \cite{Holleis2023}. The phase space where the SC gap is largest is found at the phase boundaries of SC and VPSP. As we show in Appendix \ref{DOSFX}, the strength of the SC order parameter can be tuned by the SOC by modifying the DOS near the Fermi level. 

\section{Discussion and Summary}\label{disco}

Motivated by the emergence of superconductivity in biased BBG proximate to WSe$_2$, we investigate the role of Ising SOC in promoting superconductivity. First, we provide a qualitative description for the possible origin of Ising SOC through hopping between the $p$-orbitals of graphene and the $d$-orbitals of tungsten. We then derive an interacting model based on the observation that the low-energy bands of biased BBG resemble those of a honeycomb lattice with a staggered sublattice potential, where spin-triplet superconductivity has been predicted via momentum-space Hund’s coupling \cite{spintriplet}. However, the competition between superconductivity with other symmetry-broken phases and their interplay with the van Hove singularity have yet to be explored.

By applying a Schrieffer-Wolff transformation, including further next-nearest-neighbour interactions, we find that valley-singlet spin-triplet superconductivity coexists with magnetic order near the van Hove singularity, when the Hund's coupling is sufficiently large. However, for weaker Hund's coupling, the system develops valley polarization or intervalley order, foregoing superconductivity. When Ising SOC is introduced, superconductivity is further suppressed near the van Hove singularities by a competing phase with coexisting valley and magnetic order. The enhancement of superconductivity due to Ising SOC occurs only in a small region of phase space where the enhanced density of states leads to a larger superconducting gap. This requires fine-tuning of the chemical potential. In contrast to other proposals \cite{DongZ2024,Curtis2023,Jimeno-Pozo2023,LiZ2023}, the general behaviour is that Ising SOC promotes coexisting valley and magnetic order near the van Hove singularity, suppressing superconductivity.

Based on the effective interactions derived from the Schrieffer-Wolff transformation, we speculate that the proximate WSe$_2$ monolayer effectively reduces the repulsive Coulomb interaction among doped electrons $V\sim V_0'$, which in turn weakens the attractive interactions responsible for valley and magnetic orders while enhancing those that promote superconductivity. To test this hypothesis, we propose a heterostructure of BBG with lighter 3d transition-metal monolayers that have weaker atomic SOC, to observe if superconductivity emerges. While hydrostatic pressure could be used to enhance screening, it also increases the Ising SOC itself \cite{Szentpeteri2024}, making it difficult to disentangle the two effects. Future theoretical and experimental studies are needed to understand the enhancement of superconductivity in BBG heterostructures with strong Ising SOC. Our study is limited to $\mbfq = 0$ order parameters, and finite $\mbfq \neq 0$ orders including intervalley magnetic order and FFLO superconducting states are also interesting subjects for future study.

\begin{acknowledgments}
This work was supported by the Natural Sciences and Engineering Research Council of Canada (NSERC) Discovery Grant No. 2022-04601 and the Canadian institute for Advanced Research (CIFAR). H.Y.K acknowledges support from the Canada Research Chairs Program and the Simons Emmy Noether fellowship program of the Perimeter Institute, supported by a grant from the Simons Foundation (1034867, Dittrich). Computations were performed on the Niagara supercomputer at the SciNet HPC Consortium. SciNet is funded by: the Canada Foundation for Innovation under the auspices of Compute Canada; the Government of Ontario; Ontario Research Fund - Research Excellence; and the University of Toronto.

\end{acknowledgments}

\bibliography{biblio}

\onecolumngrid
\appendix

\section{Low-energy dispersion}\label{lownrgappendix}
Here we give the expressions for the parameters $u,v$ and $w$ introduced in Eq. (\ref{lowenergyexpansion}). We have 
\begin{align*}
    u & = -\frac{9a_0^2}{8}\frac{t_\|^2}{t_\perp^2}D\left( 4-\frac{t_3^2}{D^2}\frac{t_\perp^2}{t_\|^2} \right), \\
    v & = \frac{9a_0^3}{16}\frac{t_\|^2}{t_\perp^2}D\left( 4 + 6\frac{t_3t_\perp }{D^2} - \frac{t_3^2}{D^2}\frac{t_\perp^2}{t_\|^2} \right), \\
    w & = \frac{27a_0^4}{128}\frac{t_\|^4}{t_\perp^4}D\left[ 12\left( 4+ \frac{t_\perp^2}{D^2} + 16\frac{D^2}{t_\perp^2} \right) + 4\frac{t_\perp^2}{t_\|^2}\left( 1 - 3\frac{t_3 t_\perp}{D^2} -12\frac{t_3^2}{t_\perp^2} \right) - \frac{t_\perp^4}{t_\|^4}\left( \frac{t_3^2}{D^2}+3\frac{t_3^4}{D^4} \right) \right]. \numberthis
\end{align*}
In the low-field limit, i.e. when $D\ll t_\perp,t_3,$
\begin{align*}
    u & = -\frac{9a_0^2}{8}\frac{t_\|^2}{t_\perp^2}D\left(4-\frac{t_3^2}{D^2}\frac{t_\perp^2}{t_\|^2}\right), \\
    v & = \frac{27a_0^3}{8}\frac{t_\|^2 t_3}{D t_\perp}, \\ w & = \frac{27a_0^4}{128}\frac{t_\|^4}{t_\perp^4}D\left( 12\frac{t_\perp^2}{D^2}-12\frac{t_\perp^2}{t_\|^2}\frac{t_\perp t_3}{D^2} -3\frac{t_\perp^4}{t_\|^4}\frac{t_3^4}{D^4} \right). \numberthis 
\end{align*}
In the high-field limit, that is, when $D\sim t_\perp,t_3,$
\begin{align*}
    u & = -\frac{9a_0^2t_\|^2}{2t_\perp^2}D, \\
    v & = \frac{9a_0^3}{8}\frac{t_\|^2}{t_\perp^2}D\left( 2 + 3\frac{t_3 t_\perp}{D^2}\right), \\
    w & = \frac{81a_0^4}{32}\frac{t_\|^4}{t_\perp^4}D\left(4+ \frac{t_\perp^2}{D^2} +16\frac{D^2}{t_\perp^2} \right). \numberthis
\end{align*}
In terms of $u,$ $v,$ and $w,$ the $k_y$-coordinate for the van Hove singularity along $k_x=0$ is found to be
\begin{equation}
    k_{\rm{vHS}} = \frac{-3v+\sqrt{9v^2-3wu}}{8w}>0,
\end{equation}
as measured from the $\mathbf{K}$ point. In terms of energy,
\begin{equation}
    E_{\rm{vHS}} = D + \frac{u}{32w^2}\left( 9v^2 -8uw \right)  + \frac{v}{512w^3}\left[ \left(9v^2-32wu\right)^{3/2} -27v^3 \right].
\end{equation}

\section{Non-interacting Fermi surface topology}\label{smallSOC}
\begin{figure}[h]
    \centering
    \includegraphics[scale=0.7]{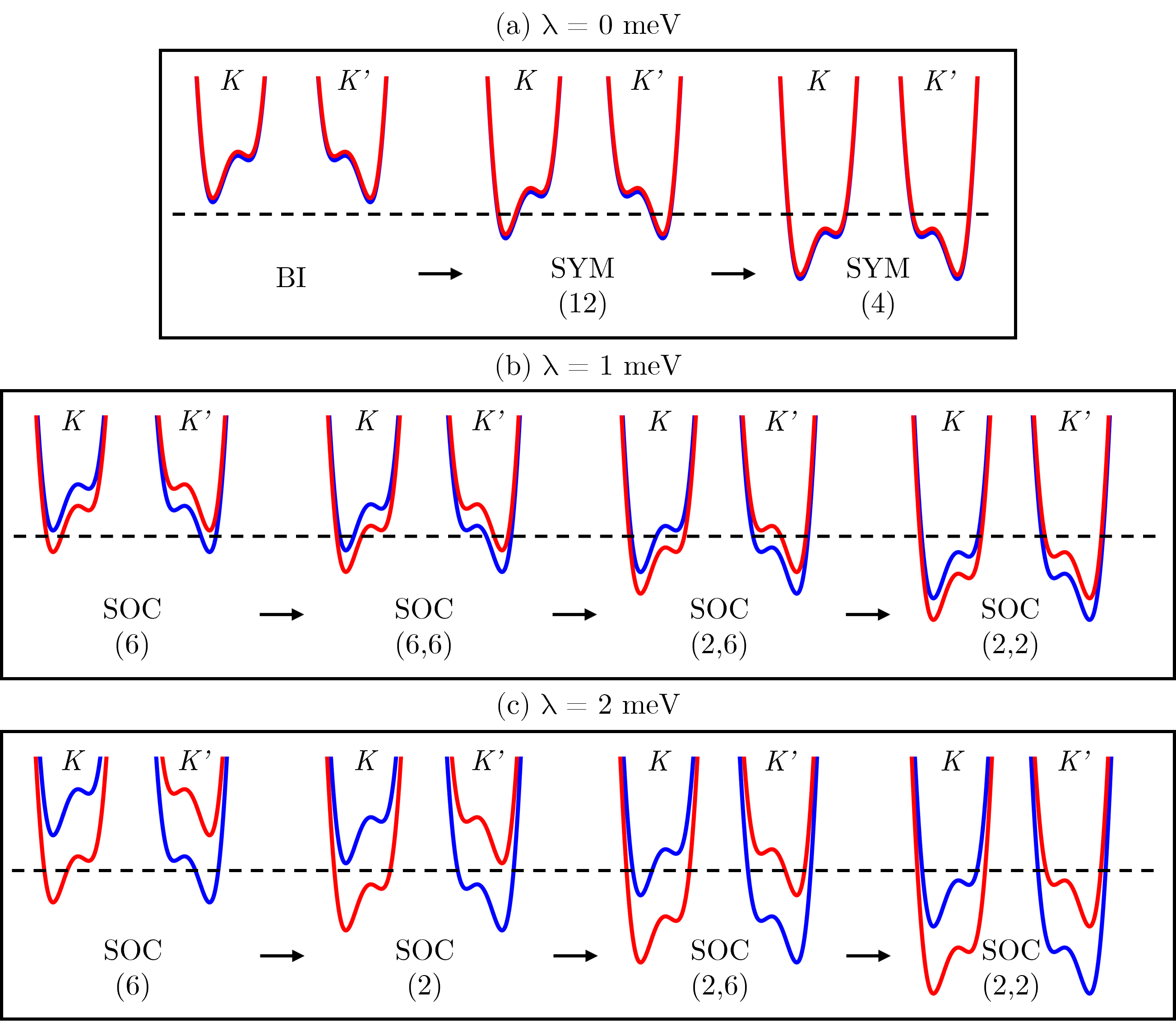}
    \caption{Non-interacting Fermi surface topology for the tight-binding model described in Sec. \ref{Theory} for (a) $\lambda=0$ meV, (b) $\lambda = 1$ meV, and (c) $\lambda=2$ meV. Momentum cuts are along $k_x=0.$ Colour denotes spin.}
    \label{topologiesAppendix}
\end{figure}

In Fig. \ref{topologiesAppendix} we show some FS topologies for the non-interacting Hamiltonian $\mathcal{H}_0 + \mathcal{H}_\lambda$ for three representative values of the Ising SOC. If the SOC splitting $2\lambda<W,$ where $W$ is the bandwidth between the band edge and the local maximum, some FS topologies are available that not possible for $2\lambda>W.$ For example, the topology SOC(6,6) possessing twelve Fermi pockets (six larger and six smaller) is available for $\lambda=1$ meV, but not for $\lambda=2$ meV.

\section{Particle-hole transformation}\label{PHT}
In this appendix, we demonstrate that our results apply to both electron doping into the conduction band and hole doping into the valence band. To do so, we show how the Hamiltonian is invariant under a particle-hole transformation combined with sublattice exchange. The Hamiltonian is
\begin{align*}
   \mathcal{H} & = t_0\sum_\sigma\sum_{\langle i,j\rangle} \left( a_{1,i,\sigma}^\dagger b_{2,j,\sigma} + h.c.\right) + D \left(\,\, \sum_{i\in A_1}n_{i} - \sum_{i\in B_2}n_{i} \right)  \\ & + U_{A}\sum_{i\in A_1} n_{i\uparrow} n_{i\downarrow} + U_{B}\sum_{i\in B_2} n_{i\uparrow} n_{i\downarrow} + V_0\sum_{\langle i,j\rangle} n_i n_j +  V_A'\sum_{\langle\langle i,j\rangle\rangle\in A_1} n_i n_j + V_B'\sum_{\langle\langle i,j\rangle\rangle\in B_2} n_i n_j, \numberthis
\end{align*}
where we have included an effective hopping $t_0$ that acts between the $A_1$ and $B_2$ sites. The NN Hubbard interaction $V_0$ acts between $A_1$ and $B_2$ sites, while $V_A'$ ($V_B'$) acts between $A_1$ and $A_1$  ($B_2$ and $B_2$) at NNN distance. This Hamiltonian is invariant under $a_{1,i,\sigma}^\dagger \rightarrow -b_{2,-i,\sigma}$ and $b_{2,i,\sigma}^\dagger\rightarrow a_{1,-i,\sigma},$ provided $U_{A}=U_{B}$ and $V_A'=V_B'.$ This can be thought of as a combined particle-hole transformation with inversion. Inversion $\mathcal{I}$ maps $A_1$ sites to $B_2$ sites, and can be written as
\begin{align*}
    \mathcal{I}^{-1}a_{1,i,\sigma}^\dagger\mathcal{I} & = -b_{2,-i,\sigma}^\dagger, \\
    \mathcal{I}^{-1}b_{2,i\sigma}^\dagger\mathcal{I} & = a_{1,-i,\sigma}^\dagger. \numberthis
\end{align*}
The minus sign takes care of the hopping term ($t_0$), and the mapping of $A_1$ sites to $B_2$ sites takes care of the sign of the on-site potential term ($D$). The particle-hole transformation can be written
\begin{align*}
    \mathcal{C}^{-1}a_{1,i,\sigma}^\dagger\mathcal{C} & = a_{1,i,\sigma}, \\
    \mathcal{C}^{-1}b_{2,i,\sigma}^\dagger\mathcal{C} & = b_{2,i,\sigma}. \numberthis
\end{align*}
The combined transformation $\mathcal{T} = \mathcal{CI}$ gives
\begin{align*}
    \mathcal{T}^{-1}a_i^\dagger\mathcal{T} & = -b_{-i}, \\
    \mathcal{T}^{-1}b_i^\dagger\mathcal{T} & = a_{-i}. \numberthis
\end{align*}
The $i\rightarrow -i$ part is not very important, as the lattice sums can simply be reindexed. Under the transformation $\mathcal{T},$ the Hamiltonian is transformed as $\mathcal{H} \rightarrow \mathcal{H}'=\mathcal{T}^{-1}\mathcal{H}\mathcal{T}$ and we obtain
\begin{align*}
   \mathcal{H}'  & = t_0\sum_\sigma\sum_{\langle i,j\rangle} \left( a_{1,i,\sigma}^\dagger b_{2,j,\sigma} + h.c.\right) + (D-U_B-3V_0-6V_B')\sum_{i\in A_1} n_{i} - (D+U_A+3V_0+6V_A')\sum_{i\in B_2} n_{i}
   \\ & + U_{B}\sum_{i\in A_1} n_{i\uparrow}n_{i\downarrow} + U_{A}\sum_{i\in B_2} n_{i\uparrow} n_{i\downarrow} + V_0\sum_{\langle i,j\rangle} n_i n_j +  V_B'\sum_{\langle\langle i,j\rangle\rangle\in A_1} n_i n_j + V_A'\sum_{\langle\langle i,j\rangle\rangle\in B_2} n_i n_j. \numberthis
\end{align*}
Up to an overall shift in the chemical potential, the transformed Hamiltonian $\mathcal{H}'$ is equivalent to $\mathcal{H}$ provided $U_A=U_B$ and $V_A'=V_B'.$ A chemical potential term $-\mu \left(\,\, \sum_{i\in A_1}n_{i} + \sum_{i\in B_2}n_{i} \right)$ can also be included, which changes sign under the prescribed transformation.

\section{Renormalized interactions}\label{renorm}
The Schrieffer-Wolff transformation outlined in Sec. \ref{Interacting} and pioneered by Ref. \cite{spintriplet}, including up to next-nearest-neighbour interactions, yields the following renormalized interaction strengths:
\begin{align*}\label{renormeq}
    \Tilde{t} & =\frac{t_0^2}{\bar{D}+2V_0'}+\frac{t_0^2}{\bar{D}+U_{A_1}+3V_0'}-\frac{2t_0^2}{\bar{D}+V_0}, \quad \gamma = \frac{t_0^2}{\bar{D}+2V_0'} - \frac{t_0^2}{\bar{D}+V_0},
    \\
    U & = U_{A_1} - \frac{6t_0^2\left(4V_0^2+V_0\bar{D} + \bar{D}^2+3V_0'\left(\bar{D}-2V_0+2V_0'\right)\right)}{\left(\bar{D}+2V_0'\right)\left(\bar{D}+V_0+V_0'\right)\left(\bar{D}+2V_0\right)} + \frac{6t_0^2}{\bar{D}+V_0+U_{A_1}},
    \\
    V & = V_0' + \frac{4t_0^2\left(V_0\left(\bar{D}-V_0\right)-V_0'\left(\bar{D}+3V_0'-4V_0\right)\right)}{\left(\bar{D}+2V_0'\right)\left(\bar{D}+V_0+V_0'\right)\left(\bar{D}+2V_0\right)} - \frac{2t_0^2\left(V_0-2V_0'\right)}{\left(\bar{D}+U_{A_1}+2V_0'\right)\left(\bar{D}+V_0+U_{A_1}\right)}, \numberthis
\end{align*}
where $\bar{D} \equiv 2D - U_{A_1} + 3V_0 -12 V_0'.$ One can verify that our model reduces to that of Ref. \cite{spintriplet} when $V_{0}'\rightarrow 0.$

As discussed in the main text, we do not find SOC to enhance the superconducting gap, but merely to pin the orientation of the $d$-vector. We speculate that, in addition to inducing SOC, the WSe$_2$ substrate also renormalizes the electron-electron interactions. Specifically, we expect the substrate to screen long-range interactions among the doped electrons in the graphene. From Eq. (\ref{renormeq}), if the NNN Coulomb repulsion $V_0'$ is decreased, $\gamma$ will increase and $V$ will decrease. The triplet interaction, when expressed in terms of the original lattice parameters, is given by $2V_T=g_1+g_2/4=9V-18\gamma.$ Thus, the effect of screening is to make the triplet interaction more attractive. We note that this line of argument does not require differential screening for the two graphene layers \cite{WSe2}.

\section{Order parameters in Nambu space}\label{NambuAppendix}
Putting together all three terms, the full MF Hamiltonian is given by ${\cal H}_{\rm MF} \equiv {\cal H}_0 + {\cal H}_{\lambda} + {\cal H}_{\rm int}^{\rm MF}$. It is convenient to express ${\cal H}_{\rm MF}$ in Nambu space by introducing three Pauli matrices, $\tau$, $\sigma$, and $\rho$ acting on the valley, spin, and particle-hole degrees o freedom, respectively. Introducing
\begin{equation}
    {\tilde \Psi}_{\bf p} = (c_{+, \mbfp,\uparrow},c_{-,\mbfp,\uparrow}, c_{+,\mbfp,\downarrow},c_{-,\mbfp,\downarrow}, c^\dagger_{+, -\mbfp,\uparrow},c^\dagger_{-, -\mbfp,\uparrow}, c^\dagger_{+,-\mbfp,\downarrow},c^\dagger_{-,-\mbfp,\downarrow})^T,
\end{equation}
${\cal H}_{\rm MF}$ is written as $\sum_{\bf p} {\tilde \Psi}_{\bf p}^\dagger H_{\bf p} {\tilde \Psi}_{\bf p}$, where
\begin{align*}
    H_{\mbfp} & = \frac{1}{2} \xi_\mbfp \rho_z \sigma_0 \tau_0 + \frac{1}{2} \varepsilon^a_\mbfp \rho_z \sigma_0 \tau_z + \frac{1}{2} \lambda \rho_z \sigma_z \tau_z \\ & + \frac{1}{2}V_P\left[ P_x\rho_z \sigma_0 \tau_x + P_y\rho_z \sigma_0 \tau_y + P_z\rho_z \sigma_0 \tau_z \right] + \frac{1}{2}V_M\left[ M_x\rho_z \sigma_x \tau_0 + M_y\rho_z \sigma_y \tau_0 + M_z\rho_z \sigma_z \tau_0 \right] \\ & +
    \frac{1}{2}V_T \left[-\Delta_x' i\rho_y \sigma_z i\tau_y + \Delta_y' \rho_x i\sigma_0 i\tau_y + \Delta_z' i\rho_y \sigma_x i\tau_y\right] +\frac{1}{2}V_T \left[-\Delta_x'' i\rho_x \sigma_z i\tau_y - \Delta_y'' \rho_y i\sigma_0 i\tau_y + \Delta_z'' i\rho_x \sigma_x i\tau_y \right].
\end{align*}
Here, $\Delta^\prime$ and $\Delta^{\prime\prime}$ are the real and imaginary part of the triplet order parameter, respectively. The factors of $1/2$ compensate the Nambu doubling. Simplifying a little bit we have
\begin{align*}
    H_{\mbfp} & = \frac{1}{2} \rho_z\left( \xi_\mbfp \sigma_0 \tau_0 + \varepsilon^a_\mbfp \sigma_0 \tau_z \right) + \frac{1}{2} \lambda \rho_z \sigma_z \tau_z + \frac{1}{2}V_P\rho_z\left( \mathbf{P}\cdot\boldsymbol{\tau} \right)\sigma_0 + \frac{1}{2}V_M\rho_z \left( \mathbf{M}\cdot\boldsymbol{\sigma} \right)\tau_0 \\ & +
    \frac{1}{2}V_T \left[-\Delta_x' i\rho_y \sigma_z + \Delta_y' \rho_x i\sigma_0 + \Delta_z' i\rho_y \sigma_x \right]\left(i\tau_y\right) +\frac{1}{2}V_T \left[-\Delta_x'' i\rho_x \sigma_z - \Delta_y'' \rho_y i\sigma_0 + \Delta_z'' i\rho_x \sigma_x \right]\left(i\tau_y\right).
\end{align*}

\section{Numerical details}\label{numericaldeets}
These calculations were performed at zero temperature on a (roughly) $70,000$ $k$-point hexagonal sampling of the Brillouin zone, at two small patches in the vicinity of the $\pm K$ points (see Fig. \ref{BZ}). The momentum cutoff is defined by the corners of the hexagonal grid, which are 1/110 the magnitude of a reciprocal lattice vector. The grid orientation was chosen to possess the same six-fold symmetry as the spin-triplet order parameter.

\section{Fermi surface topology of the ordered phases}\label{evobandtop}

\begin{figure}
    \centering
    \includegraphics[width=0.8\linewidth]{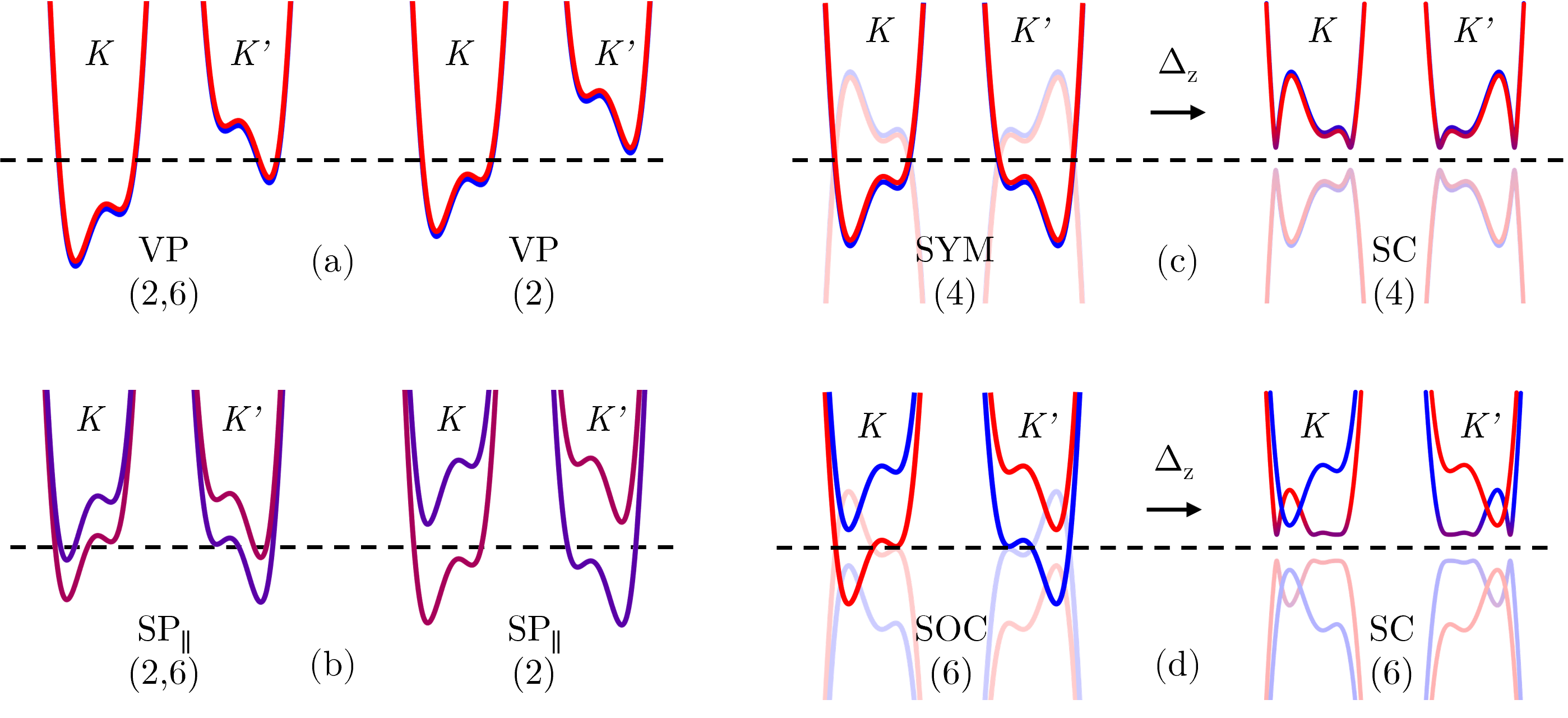}
    \caption{Fermi surface topologies for selected ordered phases: (a) Valley-polarized phases without Ising SOC. (b) Spin-polarized phases with magnetization axis oriented in-plane and nonzero SOC. (c) Superconducting transition from the symmetric normal state without Ising SOC. (d) Superconducting transition from the normal state with Ising SOC. Colour denotes spin.}
    \label{simplephases}
\end{figure}

\begin{figure}
    \centering
    \includegraphics[width=0.5\linewidth]{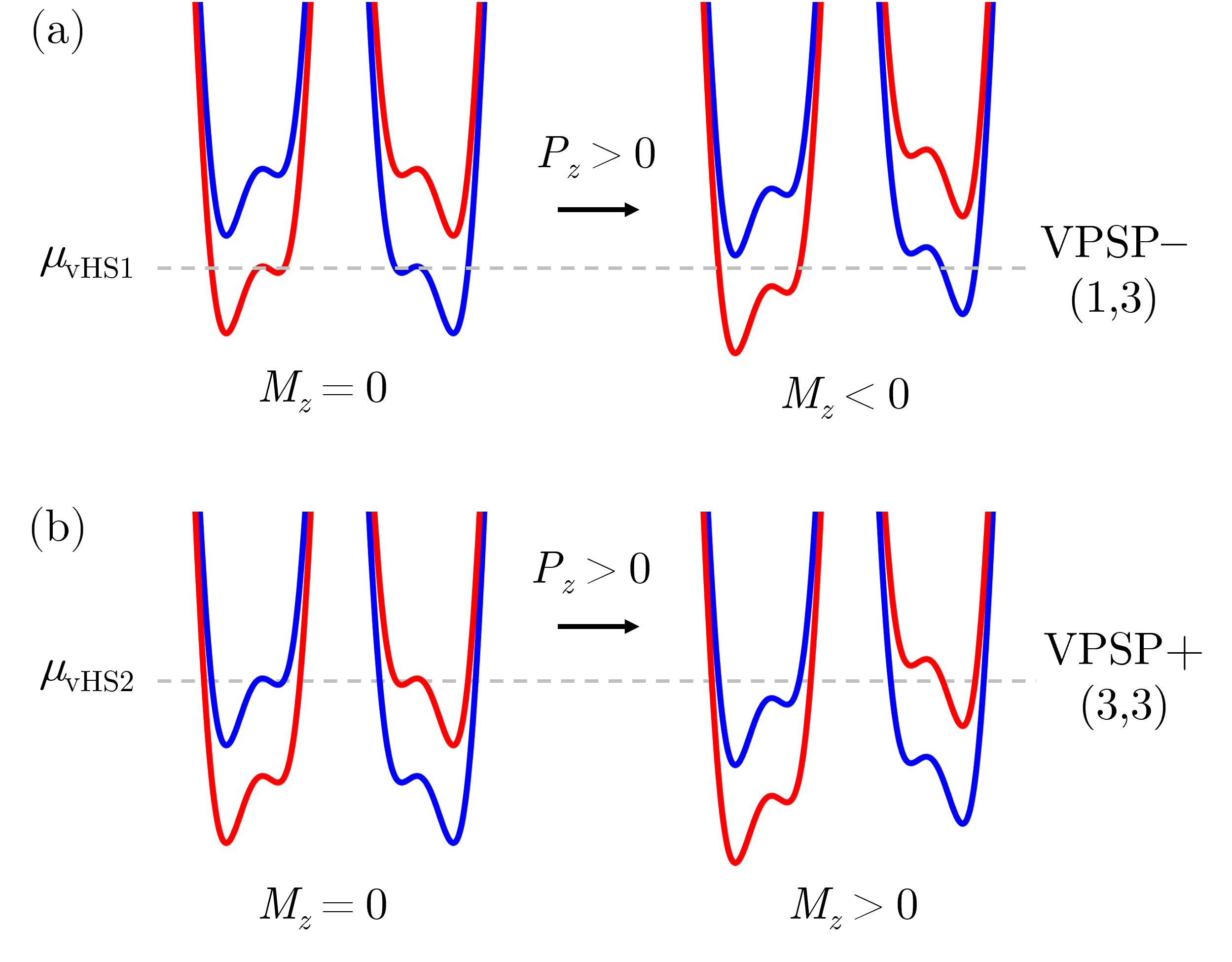}
    \caption{Sign of $P_zM_z.$ Blue and red colours correspond to spin-up and spin-down. (a) If the valley polarization develops about $\mu_{\rm vHS_1},$ the resulting phase has $P_zM_z<0,$ i.e. VPSP$-$ (the FS topology is also indicated). (b) If the valley develops about $\mu_{\rm vHS_2},$ the resulting phase has $M_zP_z>0$ i.e. VPSP$+$. The non-interacting phases are at a transition between two topologies, namely (a) SOC(6)/SOC(2) and (b) and SOC(2,6)/SOC(2,2).}
    \label{mzpzschematic}
\end{figure}

\begin{figure}
    \centering
    \includegraphics[width=0.8\linewidth]{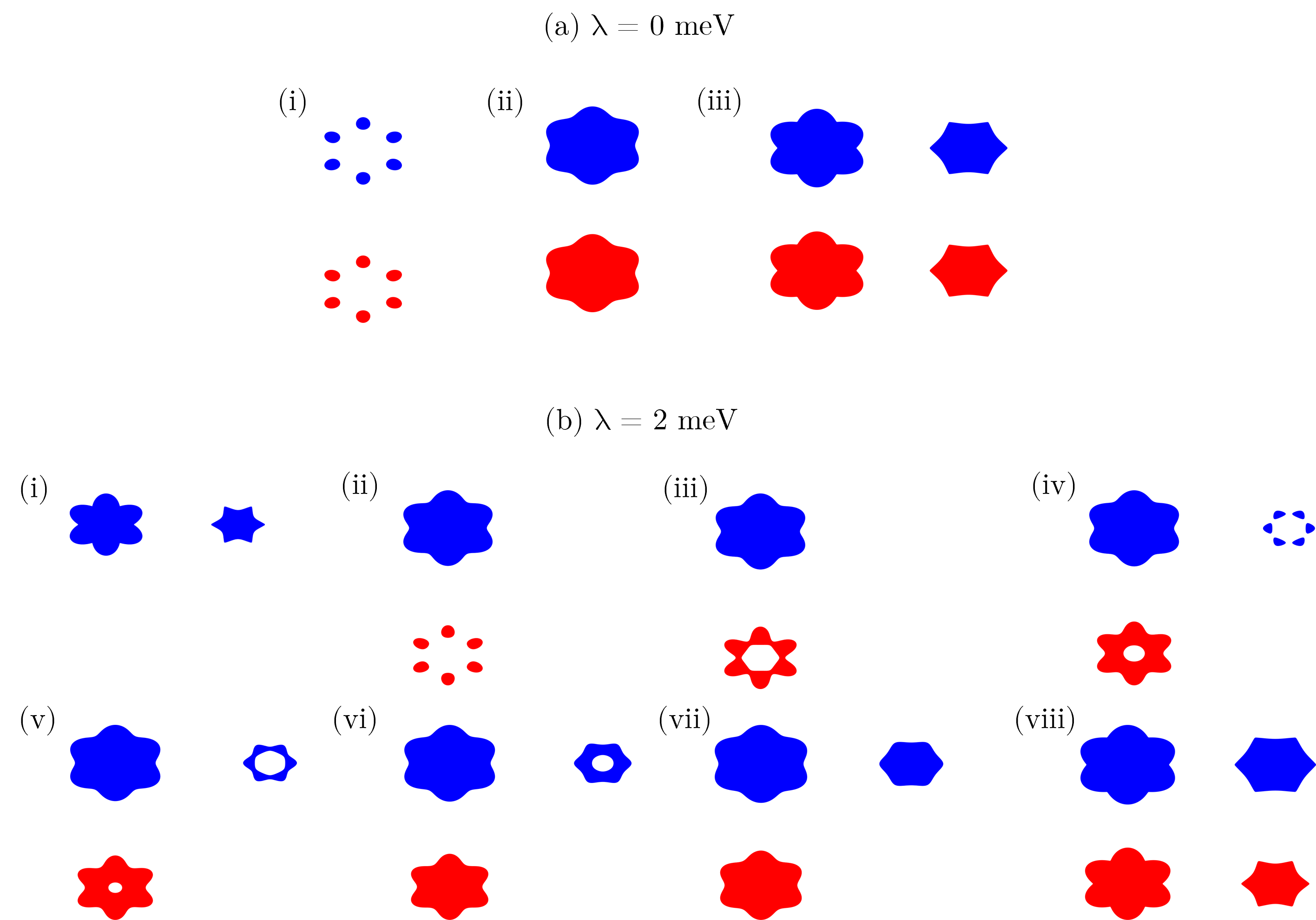}
    \caption{FS topologies for IVO. (a) $\lambda=0$ and the chemical potential increases through (i)--(iii). Cases (i) and (iii) are actually disordered phases with $P_x=P_y\equiv P_\|=0,$ but help to demonstrate the hybridization. Case (ii) has finite $P_\|.$ (b) $\lambda\neq 0$ and the chemical potential increases through (i)--(viii). Cases (i) and (viii) are disordered phases with $P_\|=0.$ All other cases have finite $P_\|.$ As can be seen, the FS topology changes rapidly within the IVO phase as the chemical potential is increased. Colour denotes spin, but the relative positions of the Fermi surfaces are arbitrary as valley index is no longer a good quantum number.}
    \label{ivophases}
\end{figure}

Here we present the Fermi surface topologies for a few select symmetry-breaking phases. In Fig. \ref{simplephases}, we examine valley-polarized (VP), spin-polarized (SP), and superconducting (SC) FS topologies. 
Next we examine the $\rm{VPSP}\pm$ phases, which are distinguished by the sign of the product of the VP and SP order parameters, $P_zM_z.$ As shown in Fig. \ref{mzpzschematic}, the sign of $P_zM_z$ depends on whether the chemical potential is tuned to the upper or lower vHS, which determines which pair of SOC-split bands are involved in the Pomeranchuk transition. The spins of the active bands depend on the sign of $\lambda,$ so we assume fixed signs of $\lambda>0$ and $P_z>0.$
Finally, in Fig. \ref{ivophases}, we plot the FS topologies for intervalley order (IVO). Specifcally, we examine the topologies that develop in case (a) from Fig. \ref{MFplots} in the main text. The topologies associated with IVO involve hybridization between the $K$ and $K'$ valleys, and are therefore more complicated than those we have examined so far. Additionally, the hybridization prevents us from cleanly separating the bands by their valley index, although $S_z$ remains a good quantum number.

\section{DOS effects of SOC}\label{DOSFX}
\begin{figure}
    \centering
    \includegraphics[width=0.6\linewidth]{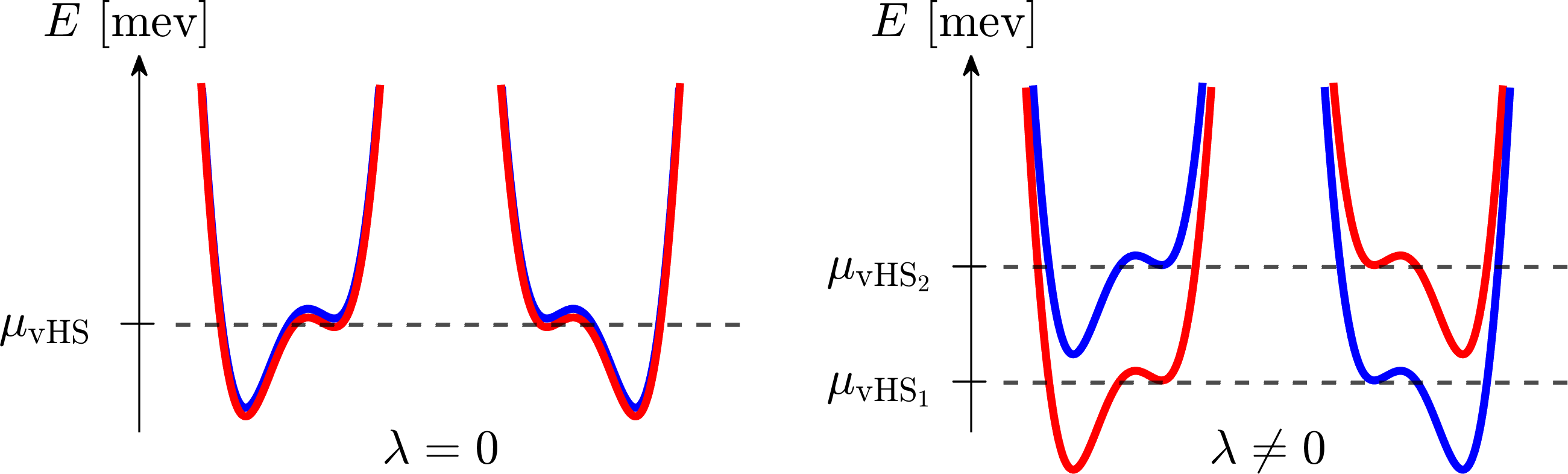}
    \caption{Degradation of the pairing surface in the presence of SOC. Left panel: $\lambda=0$ and the chemical potential is aligned with the van Hove singularity. Right panel: $\lambda\neq 0$ and the chemical potential can only be aligned with one of two non-degenerate van Hove singularities.}
    \label{appendixschematic}
\end{figure}

In this appendix, we examine the effect of Ising SOC on the superconducting gap in the absence of any other order parameters. In particular, we focus on how the SOC tunes the DOS near the Fermi level. When $\lambda=0,$ the superconducting gap is maximal when the chemical potential $\mu$ is aligned with the four-fold degenerate van Hove singularity at $\mu_{\rm vHS} = D$. Nonzero $\lambda$ causes the bands to polarize according to their spin, with opposite sign in $K$ and $K'.$ As a result, $\lambda$ displaces the vHS to $\mu_{\rm vHS_1} = D-\lambda$ and $\mu_{\rm vHS_2} = D + \lambda$ (see Fig. \ref{appendixschematic}). In this case, $\Delta_z$ wil be maximal if the chemical potential is aligned with $\mu_{\rm vHS_1}$ or $\mu_{\rm vHS_2}$. Aligning with $\mu_{\rm vHS_1}$ is less preferable, as pairing can only occur between the lower bands. Aligning with $\mu_{\rm vHS_2}$ allows for pairing in all bands, but will be weaker than the $\lambda=0$ case because the Fermi surface cannot be aligned with both vHSs simultaneously.

\begin{figure}
    \centering
    \includegraphics[width=0.8\linewidth]{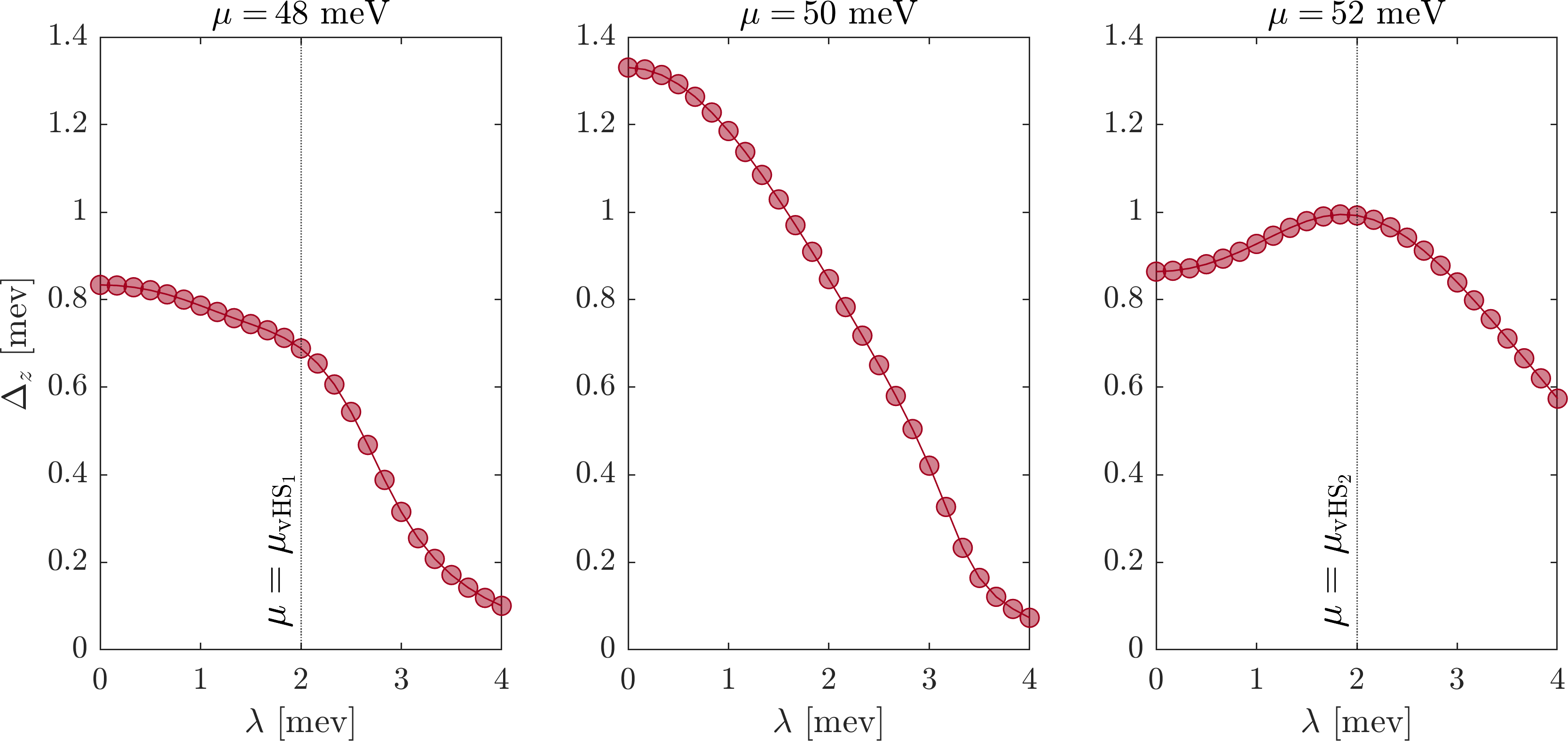}
    \caption{Magnitude of the superconducting gap as a function of $\lambda$ for three choices of chemical potential. In (b), the chemical potential is set to $\mu_{\rm vHS}.$ In (a) and (c), the chemical potential is placed 2 meV below and above $\mu_{\rm vHS},$ respectively. When $\lambda=2$ meV the chemical potential is re-aligned with either $\mu_{\rm vHS_1}$ or $\mu_{\rm vHS_2}.$}
    \label{appendixlambda}
\end{figure}

In Fig. \ref{appendixlambda}, we plot the superconducting gap for the triplet in the absence of any other interactions for three choices of $\mu.$ Note that when $\lambda\neq 0,$ the $\Delta_z$ component is selected over $\Delta_x$ and $\Delta_y.$ In the middle panel, the chemical potential is set to $\mu_{\rm vHS} =50$ meV. As $\lambda$ increases, $\mu_{\rm vHS}$ splits into $\mu_{\rm vHS_1}$ and $\mu_{\rm vHS_2}$ but the chemical potential is held fixed. The pairing surface therefore degrades as $\mu$ is no longer aligned with a vHS. In the rightmost panel, the chemical potential is placed above $\mu_{\rm vHS}.$ As $\lambda$ is increased, $\mu_{\rm vHS_2}$ passes through the Fermi level and the SC gap experiences a maximum. In the leftmost panel, the chemical potential is placed below $\mu_{\rm vHS}.$ As $\lambda$ is increased, $\mu_{\rm vHS_1}$ passes through the Fermi level but no maximum is observed as pairing can only occur in the lower set of bands (see Fig. \ref{appendixschematic}).

\end{document}